\documentclass[a4paper,usenatbib]{mnras}
\usepackage{graphicx}
\usepackage[fleqn]{amsmath}
\usepackage[T1]{fontenc}
\usepackage{aecompl}
\usepackage{times}
\usepackage{ae,aecompl}
\usepackage{amssymb,amsfonts,hyperref,color}

\newcommand\be{\begin{equation}}
\newcommand\en{\end{equation}}

\title[Misaligned Planets]{Inclined Massive Planets in a Protoplanetary Disc: Gap Opening, Disc Breaking, and Observational Signatures }

\author[Zhu]{Zhaohuan Zhu$^{1}$\thanks{E-mail: zhaohuan.zhu@unlv.edu}\\
$^{1}$Department of Physics and Astronomy, University of Nevada, Las Vegas, 
                4505 S. Maryland Pkwy, Las Vegas, NV, 89154, USA\\}
\date{In original form \today}

\begin{document}
\label{firstpage}
\pagerange{\pageref{firstpage}--\pageref{lastpage}} \pubyear{2018}
\maketitle

\begin{abstract}
We carry out
three-dimensional hydrodynamical simulations to study  planet-disc interactions for inclined high mass planets, focusing on the disc's secular evolution induced by the planet. We find that, when the
planet is massive enough and the induced gap is deep enough, the disc inside the planet's orbit breaks from the outer disc.
The inner and outer discs precess around the system's total angular momentum vector independently at different precession rates, which causes significant disc misalignment. We derive the analytical formulae, which are also verified numerically, for: 1) the relationship between the planet mass and the depth/width of the induced gap, 2) 
the migration and inclination damping rates for massive inclined planets, and 3) the condition under which the inner
and outer discs can break and undergo differential precession. 
Then, we  carry out Monte-Carlo radiative transfer
calculations for the simulated broken discs. Both disc shadowing in near-IR images and gas kinematics probed by molecular lines (e.g. from ALMA) can reveal the misaligned inner disc. The relationship between the rotation rate of the disc shadow and the precession rate of the inner disc is also provided. Using our disc breaking condition, we conclude that the
disc shadowing due to misaligned discs should be accompanied by deep gaseous gaps (e.g. in Pre/Transitional discs). This scenario naturally explains both the disc shadowing and deep gaps in several systems (e.g. HD 100453, DoAr 44, AA Tau,  HD 143006) and these systems should be the prime targets for searching young massive planets ($>M_J$) in discs.

\end{abstract}

\begin{keywords}
planet-disc interaction --- protoplanetary discs --- accretion, accretion discs --- hydrodynamics --- radiative transfer --- planet and satellites: detection 
\end{keywords}

\section{Introduction}
The planet-disc interaction theory has been developed over decades \citep{GoldreichTremaine1979,Tanaka2002, KleyNelson12, Baruteau2014}.  But most studies focus on planets that are coplanar with the disc. This is due to both the fact that planets in our Solar Systems are coplanar and the simplicity of the coplanar case where the vertically averaged two-dimensional equations can be derived and 2-D numerical simulations are normally sufficient \citep{muller2012,FungChiang2016}. 

However, many discovered exoplanets
have their orbital angular momentum vector misaligned with the stellar spin axis (see review by \citealt{WinnFabrycky2015}). Especially for Hot Jupiters ($M_{p}>$0.3 $M_{J}$ and $P<10$ days), the misalignment spans the entire range from $0^o$ to $180^o$, revealed by  measurements of the
Rossiter-McLaughlin effect  \citep{Albrecht2012}.  The misalignment can be due to  1)  processes through which the planet moves away from the disc midplane where it formed, such as planet-planet scattering (e.g. \citealt{Chatterjee2008,JuricTremaine2008}), Kozai-Lidov oscillations (e.g. \citealt{WuMurray2003,Petrovich2015}),  and secular chaos \citep{WuLithwick2011}, or 2) the primordial misalignment  of the natal disc with respect to the stellar spin during the disc evolution,  such as disc formation in a turbulent environment \citep{Bate2010}, magnetic torque from the star \citep{Lai2011}, or the gravitational torque from a misaligned companion star \citep{Batygin2012, Martin2016}. In either scenario, as long as the planet is misaligned with the disc, understanding how the misaligned planet can interact with the disc is crucial for studying the planet's orbital evolution afterwards. 

For an inclined low-mass planet which does not induce gaps in the disc, its interaction with the disc has been studied
analytically in both the small inclination limit using the linear perturbation theory \citep{TanakaWard2004} and the large inclination limit with the dynamical friction theory \citep{Rein2012}. Numerically, \cite{Cresswell2007} and \cite{BitschKley2011} have confirmed that low mass planets with small inclinations are undergoing exponential inclination decay, consistent with the linear theory. However, their measurements for moderately inclined planets are not consistent with the dynamical friction theory. This disagreement  is likely due to the narrow range of the planet inclination explored by these simulations, as pointed out by \cite{Arzamasskiy2018}. \cite{Arzamasskiy2018} have measured the planet's migration and inclination damping rate for planets with inclination from 0$^o$ to 180$^o$, and found a good agreement between the simulation results and the analytical theory.  

While the inclined low-mass planets are relatively well studied, the situation is less clear for high mass planets which can induce gaps. \cite{MarzariNelson2009} have found that
both the eccentricity and inclination of giant planets damp very quickly, on the timescale of hundreds of orbits. On the other hand, \cite{Xiang-GruessPapaloizou2013} and \cite{Bitsch2013} have found that 
the damping rates are reduced when a gap is opened in the disc. However, 
it is not quantitatively understood how migration and inclination damping rates depend on the gap depth and how the gap depth depends on the planet mass. Furthermore, there could be long-term secular interactions between the planet and the disc \citep{LubowOgilvie2001}.

While most studies focus on the planet's orbital evolution in discs, the planet can also affect the disc structure which has barely been studied.
\cite{Xiang-GruessPapaloizou2013} have found that the presence of a massive planet can cause a warp in the protoplanetary disc. However, the warp in the disc due to even a 6 $M_J$ planet only has $\lesssim$ 20$^o$ misalignment, which is not enough to explain some of the observational signatures of disc misalignment.

Observationally, some protoplanetary discs show dark spikes on their near-IR scattered light images. The dark spikes are best explained as the shadows cast by a misaligned inner disc blocking the stellar light \citep{Marino2015, Stolker2016, Benisty2017, Debes2017, Long2017, Min2017, Casassus2018, Benisty2018}. The inferred relative inclination between the inner and outer discs can be quite large: $\sim70^o$ for HD 142527 \citep{Marino2015}, $\sim72^o$ for HD 100453 \citep{Benisty2017}, $\sim30^o$ for DoAr 44 \citep{Casassus2018}, and $\sim30^o$ for HD 143006 \citep{Benisty2018}.
Furthermore, some dark spikes vary with time, indicating the change of the inner disc on short timescales \citep{Stolker2016, Debes2017}. At much longer wavelengths, ALMA detect twisted flow patterns in the disc, directly suggesting a warped inner disc \citep{Rosenfeld2014,Pineda2014,Casassus2015, Brinch2016, Loomis2017, Walsh2017}. Finally,
the optical/near-IR light curves of some YSOs show dimming events, indicating blocking the stellar light by a warped inner disc \citep{Alencar2010,Bouvier2013, LodatoFacchini2013, Cody2014, Facchini2016, Bodman2017, Schneider2018}. 

To explain these observations,  \cite{Facchini2014, Juhasz2017, Facchini2018, Price2018} propose that compact binaries can warp and break the circumbinary discs. These circumbinary disc simulations are imported into the radiative transfer code to generate the synthetic near-IR scattered light images and (sub-)mm molecular line channel maps. 
 
In this paper,  extending our previous study on planet-disc interactions for low mass planets on inclined orbits \citep{Arzamasskiy2018}, we use both the analytical theory and 3-D hydrodynamical simulations to study how a massive misaligned planet can interact with the disc. Different from previous simulations, we find that, when the
planet mass is large enough and the induced gap is deep enough, the disc inside the planet's orbit breaks from the outer disc. Then we use Monte-Carlo radiative transfer simulations to calculate the observational signatures of such misaligned discs. 
The analytical theory for gap depth/width, planet migration, and disc breaking is presented in \S 2. The numerical confirmation is presented in \S 3 and \S 4. The observational signatures are shown in \S 5. After a short discussion in \S 6, the paper is concluded in \S 7.

\section{Theoretical Framework}
The gap-opening process by planets has been studied extensively due to its importance in reducing the planet migration rate \citep{LinPapaloizou1986} and  explaining recent observations of discs with gaps and cavities (e.g. \citealt{Espaillat2014}). The quantitative relationships between  
the gap shape, migration rate, and the planet and disc properties have been worked out in great detail
 \citep{ Duffell2014, Fung2014,  Kanagawa2015, Kanagawa2016, GinzburgSari2018, DurmannKley2015}.
However, these studies only focus on  planets that are coplanar with the disc.  
Numerical simulations by
\cite{Xiang-GruessPapaloizou2013} and \cite{Bitsch2013} have indicated that misaligned massive planets can behave very differently from coplanar planets. 
For example, a gap can be induced by a misaligned planet and the gap is shallower than the gap in the coplanar case. After a gap is induced,
the planet's inclination damping rate is reduced and a warp can develop throughout the disc. 

In this work, we try to establish a theoretical framework for understanding the interaction between a misaligned high mass planet and the protoplanetary disc. We want to quantitatively answer following questions:

1) {\it How are the depth and width of the gap determined by the planet mass, the planet inclination, and the disc properties ?}

2) {\it How does the gap opening process affect the planet migration rate and inclination damping rate?}

3) {\it How does the gap opening process affect the disc evolution? How
deep does the induced gap have to be for the inner disc to break from the outer disc and undergo differential precession? }

\subsection{Gap Depth and Width}
We first review the relationships between the planet mass and the gap profile (depth and width) for coplanar planets and then generalize them for inclined planets. 

A coplanar planet excites density
waves at Lindblad resonances. 
On each side of the planet, the total
amount of angular momentum that is carried away by these density waves is derived by \cite{GoldreichTremaine1979}:
\begin{equation}
T_{wave}\sim q^2\left(\frac{H}{R}\right)_{R_p}^{-3}\Sigma_{0}\Omega_p^2 R_p^4\,, \label{eq:GT79}
\end{equation}
where $\Omega_p$ is the disc orbital frequency at the planet position ($R_p$),  $H$ is the disc scale height, $\Sigma_{0}$ is the unperturbed gas surface density at the planet position, and $q=M_{p}/M_{*}$ is the mass
ratio between the companion and the central star.
Later, we also use $h$ to represent $H/R$.
Most of these waves are excited at radii with a distance of $H$ away from the planet, so-called torque cut-off \citep{GoldreichTremaine1980}.
For a planet within a gap, 
the angular momentum of the excited waves can be calculated similarly but $\Sigma_{0}$ in Equation \ref{eq:GT79} is replaced with the surface density within the gap $\Sigma_{gap}$ \citep{Fung2014}:
\begin{equation}
T_{wave}\sim q^2\left(\frac{H}{R}\right)_{R_p}^{-3}\Sigma_{gap}\Omega_p^2 R_p^4\,. \label{eq:tw}
\end{equation}
A more accurate derivation requires calculating the 
torque self-consistently considering the gap profile \citep{GinzburgSari2018}. The induced density waves will propagate in the disc carrying a constant angular momentum flux.
When the density waves
steepen into shocks \citep{GoodmanRafikov2001}, this angular momentum will be deposited to the disc, leading
to gap opening. 

In a viscous disc, a steady gap profile can be maintained when the angular momentum deposition
into the disc balances the viscous stress \citep{Fung2014}:
\begin{equation}
T_{\nu}\sim \Sigma_0 \nu\Omega_p R_p^2\,,\label{eq:tnu}
\end{equation}
where $\nu=\alpha H^2\Omega$ is the kinematic viscosity. 
Equating equations \ref{eq:tw} and \ref{eq:tnu}, \cite{Fung2014} have shown that
\begin{equation}
\frac{\Sigma_{gap}}{\Sigma_0}\sim \alpha h^5q^{-2} \,.
\end{equation}

With a more detailed torque and angular momentum flux calculation, 
\cite{Kanagawa2015} has refined the gap-depth relationship:
\begin{equation}
    \frac{\Sigma_{gap}}{\Sigma_{0}}=\frac{1}{1+0.04 K}\,,\label{eq:gapdepth}
\end{equation}
where 
\begin{equation}
    K=\alpha^{-1}h^{-5}q^2 \,,\label{eq:Kpara}
\end{equation}
which reproduces gaps from numerical simulations extremely well. 

Regarding the gap width, \cite{Kanagawa2016} have found an empirical relationship
\begin{equation}
    \frac{\Delta_{gap}}{R_{p}}=0.41 K'^{1/4}\,,\label{eq:gapwidth}
\end{equation}
where
\begin{equation}
    K'=K h^2=\alpha^{-1}h^{-3}q^2 \,.\label{eq:Kprimepara}
\end{equation}
where $\Delta_{gap}$ is the radial extend of the gap at $\Sigma=\Sigma_{0}/2$. 

For an inclined planet, we can simply extend these relationships using an averaged planet potential. 
Assuming the inclined planet's orbital momentum vector is at an angle of $i_p$ from the disc's angular momentum vector, we first calculate the projection of the planet's position onto the disc. Then, 
we  calculate the potential experienced by the disc element that is at 
a radial distance of $H$ away from this projected position. Finally, we average the derived potential 
over one planetary orbital period to get an effective planet potential onto the disc
\begin{equation}
\Phi=\frac{\int \frac{GM_{p}}{\left((R_p {\rm sin}\phi {\rm sin} i_p)^2+H^2\right)^{1/2}} d\phi}{\int d\phi}\,,
\end{equation}
where $(R_p {\rm sin}\phi {\rm sin} i_p)^2+H^2)^{1/2}$ is the distance from the planet to the position in the disc that is one scale height away from the planet's projection onto the disc, and
$\phi$ is the angle between the vector to the planet and the line of nodes between the disc plane and the planet's orbital plane. 
When $i_p$=0, this equation is $\Phi=GM_{p}/H$. 
Thus, we can generalize Equations \ref{eq:Kpara} and \ref{eq:Kprimepara} using the effective potential and replacing $q$ in these equations with
\begin{equation}
   q \rightarrow q \frac{\int \frac{1}{\sqrt{(R_p {\rm sin}\phi {\rm sin}(i_p)/H)^2+1}}d\phi}{\int d\phi}\,.\label{eq:qinclined}
\end{equation}
The integration can be simplified as 
\begin{equation}
    q \rightarrow q \frac{2}{\pi \sqrt{n+1} }\mathcal{K}(\frac{n}{n+1})\label{eq:newq}
\end{equation}
where $n$ is $(R_p{\rm sin}(i_p)/H)^2$ and  $\mathcal{K}(m)$ is the complete elliptic integral of the first kind ($m$=$k^2$ and $k$ is the elliptic modulus). 

Thus, the gap depth and width for the inclined planet are
\begin{eqnarray}
    \frac{\Sigma_{gap}}{\Sigma_{0}}=\frac{1}{1+0.04 K}\,, \nonumber\\
    \frac{\Delta_{gap}}{R_{p}}=0.41 K'^{1/4}\,, \label{eq:depthwidth}
\end{eqnarray}
with the new 
\begin{eqnarray}
    K&=&\alpha^{-1} h^{-5} q^2\frac{4}{\pi^2 (n+1) }\left[\mathcal{K}(\frac{n}{n+1})\right]^2\nonumber\\
    K'&=&K h^2\,.\label{eq:newK}
\end{eqnarray}

In the derivation above, we have ignored the effect of dynamical friction. Dynamical friction is important for planet migration (\S 2.2) but may not be important for gap opening (at least for moderately inclined planets) due to two reasons: 1) dynamical friction is a local effect which only occurs at the place where the planet moves through the disc, 2) the angular momentum exchange between the planet and the Mach cone left behind the planet is small compared with the one-side wave torque due to the Lindblad resonances. On the other hand, the detailed calculation on the gap opening due to dynamical friction still needs to be carried out in future and may be important for highly inclined planets.

\subsection{Migration and Inclination Damping For Misaligned Planets}
\cite{Arzamasskiy2018} found that the migration and the inclination damping rates for inclined low-mass planets
can be described by the linear theory when the planet is mildly inclined \citep{Tanaka2002, TanakaWard2004}  or by the dynamical friction theory when the planet is more inclined \citep{Rein2012}. 
The resulting migration rate can be described as the minimum between these two rates:
\begin{equation}
-\frac{\left<\partial_{t}R_{p}\right>}{R_{p}}={\rm min} \begin{cases}
	(2.7+1.1\alpha_s)\cdot t_{mig}^{-1}\\
	8.8 \cdot h^2 \cdot{\rm sin}^{-1}(i_p/2)\cdot {\rm sin}^{-1}(i_p)  \cdot t_{mig}^{-1}
	\end{cases}\label{eq:ptr}
\end{equation}
where
\begin{equation}
t_{mig}=\Omega_{p}^{-1}h^2 q^{-1}\left(\frac{\Sigma_{0}R_{p}^2}{M_{*}}\right)^{-1}\,\label{eq:tmig}
\end{equation}
is the migration timescale,
and $\alpha_{s}$ is the radial slope of the surface density profile ($\Sigma\propto R^{-\alpha_s}$).

The inclination
damping rate can also be expressed as the minimum between these two limits: 
\begin{equation}
-\frac{\left<\partial_{t}i_p\right>}{i_p}={\rm min} \begin{cases}
	0.544 \cdot t_{inc}^{-1}\\
	1.46 \cdot h^4 \cdot{\rm sin}^{-3}(i_p/2)\cdot i_p^{-1} \cdot t_{inc}^{-1}
	\end{cases}\label{eq:ptpi}
\end{equation}
where 
\begin{equation}
t_{inc}=\Omega_{p}^{-1}h^4 q^{-1}\left(\frac{\Sigma_{0}R_{p}^2}{M_{*}}\right)^{-1}\label{eq:tinc}
\end{equation}
is the inclination damping timescale. So the inclination damping timescale is shorter than the migration timescale
by a factor of $h^{-2}$. Both the migration and inclination damping timescales are shorter
for a more massive planet in a more massive and thinner disc. 

On the other hand, for a massive planet which can induce a gap in the disc, 
the migration and inclination damping timescales should become longer
since $\Sigma_{0}$ should be replaced by $\Sigma_{gap}$ considering that both wave launching and dynamical friction are proportional to the local disc surface density around the planet. Thus, we need to plug the new
gap depth (Equations \ref{eq:depthwidth} and \ref{eq:newK})
into Equations \ref{eq:tmig} and \ref{eq:tinc} to get the migration and inclination damping rates. 
The migration rate is now
\begin{equation}
-\frac{\left<\partial_{t}R_{p}\right>}{R_{p}}={\rm min} \begin{cases}
	(2.7+1.1\alpha_s)\cdot t_{mig}^{-1}\\
	8.8 \cdot h^2 \cdot{\rm sin}^{-1}(i_p/2)\cdot {\rm sin}^{-1}(i_p)  \cdot t_{mig}^{-1}
	\end{cases}\nonumber
\end{equation}
with
\begin{equation}
t_{mig}=\Omega_{p}^{-1}h^2 q^{-1}\left(\frac{\Sigma_{0}\left(\frac{\Sigma_{gap}}{\Sigma_{0}}\right)^{qs}R_{p}^2}{M_{*}}\right)^{-1}\,.\label{eq:tmiggap}
\end{equation}

And the inclination damping rate is now
\begin{equation}
-\frac{\left<\partial_{t}i_p\right>}{i_p}={\rm min} \begin{cases}
	0.544 \cdot t_{inc}^{-1}\\
	1.46 \cdot h^4 \cdot{\rm sin}^{-3}(i_p/2)\cdot i_p^{-1} \cdot t_{inc}^{-1}
	\end{cases}\nonumber
\end{equation}
with 
\begin{equation}
t_{inc}=\Omega_{p}^{-1}h^4 q^{-1}\left(\frac{\Sigma_{0}\left(\frac{\Sigma_{gap}}{\Sigma_{0}}\right)^{qs}R_{p}^2}{M_{*}}\right)^{-1}\,.\label{eq:tincgap}
\end{equation}
Here we add a free parameter $qs$ to represent the sensitivity of the planet migration and damping rates with respect to the gap depth.
If the rates are proportional to $\Sigma_{gap}$, $qs$ is 1. 
However, since a misaligned planet can also interact
with disc material outside the center of the gap, we expect $qs<1$. Later, by fitting results from numerical simulations,
we empirically derive $qs\sim 2/3$.  On the other hand, this fudge factor $qs$ is highly uncertain and may not apply to
a large disc/planet parameter range. 

\subsection{Inner Disc Precession And The Disc Breaking Condition}
As will be shown by numerical simulations in \S 4, when a deep gap is induced, the inner disc can lose connection 
with the outer disc so that the disc can break. The inner and outer discs will then precess at different
rates driven by the misaligned planet, leading to a large disc misalignment.

To understand the disc breaking, we use the equations
for small amplitude warping disturbances in a nearly inviscid disc \citep{LubowOgilvie2000} : 
\begin{eqnarray}
\Sigma R^{2}\Omega\frac{\partial {\bf l}}{\partial t}=\frac{1}{R}\frac{\partial {\bf G}}{\partial R}+{\bf T}\,\label{eq:tilt1}\\
\frac{\partial {\bf G}}{\partial t}-\left(\frac{\Omega^2-\kappa^2}{2\Omega}\right) {\bf l} \times {\bf G}+\alpha \Omega {\bf G}=\frac{PR^3\Omega}{4}\frac{\partial {\bf l}}{\partial R}\label{eq:tilt2}\,,
\end{eqnarray}
where ${\bf l}(R, t)$ is the tilt vector at  radius $R$. It is defined as 
a unit vector parallel to the angular momentum vector of each annulus in the disc. 
Equation \ref{eq:tilt1} expresses the horizontal components of the angular momentum conservation. 
2$\pi{\bf G(R,t)}$ is the horizontal internal torque in the disc, and ${\bf T(R,t)}$ is the horizontal torque density. 
Equation \ref{eq:tilt2} determines the evolution of the internal torque {\bf G}. The internal torque is affected by the shearing epicyclic motions,
radial pressure gradients, and viscous decay described by the $\alpha$ viscosity parameter.
We can integrate
Equation \ref{eq:tilt1} with $R$ to derive the precession rate of the inner disc:
\begin{equation}
\int \Sigma R^{2}\Omega\frac{\partial {\bf l}}{\partial t}RdRd\phi=\left.\int {\bf G}\right\vert_{R_{in}}^{R_{out}}d\phi+\int {\bf T}RdRd\phi\,.\label{eq:tilt21}\\
\end{equation}

For a planet in the disc, its tidal torque consists of m=0 and m=2 components \citep{Bate2002}. 
Since the m=2 component only causes oscillation of the precession rate, we  focus on the m=0 component. 
The m=0 component is from an axisymmetric potential, and can be written as
\begin{equation}
T(R)=-\frac{3}{2}\frac{GM_p \Sigma R^2 }{R_p^3}{\rm sin}(i_p) {\rm cos}(i_p) {\rm sin}^2\phi \label{eq:Torque}
\end{equation}
where $i_p$ is the angle between the disc's rotation axis and the planet's angular momentum vector. 
$\phi$ is the azimuthal angle in the disc starting from the line of nodes between the disc plane and the planet's orbital plane.

If the disc breaking occurs at $R_p$ so that the inner disc
precesses on its own, the precession rate for the inner disc can be calculated using the torque from Equation \ref{eq:Torque}.
Assuming that the whole inner disc precesses as a rigid body at a rate of $\omega_{rb}$,
$\partial {\bf l}/\partial t$ in Equation \ref{eq:tilt21} is then sin$i \times \omega_{rb}$. 
The internal torque term (the $G$ term) in Equation \ref{eq:tilt21} can also be dropped due to the disc breaking.
Plugging Equation \ref{eq:Torque} into Equation \ref{eq:tilt21},
we can derive
\begin{equation}
\omega_{rb}=-\frac{3}{4}\frac{GM_{p}}{R_p^3}{\rm cos}(i_p) \frac{\int \Sigma R^3 dR}{\int \Sigma R^3\Omega dR}\,.
\end{equation}
If the disc surface density follows $\Sigma\propto R^{-\alpha_s}$ from $R=0$ to $R_d$ with $R_{d}$ being the size
of the inner disc, we have
\begin{eqnarray}
\omega_{rb}&=&-\frac{3}{4}\frac{GM_{p}}{R_p^3}{\rm cos}(i_p)\frac{1/(4-\alpha_s)\Sigma_{d}R_{d}^4}{1/(2.5-\alpha_s)\Sigma_{d}R_{d}^4\Omega_{d}}\nonumber\\
&=&-\frac{3 GM_{p}}{4 R_p^3\Omega_{d}}\frac{2.5-\alpha_s}{4-\alpha_s}{\rm cos}(i_p)\,,
\end{eqnarray}
where $\Sigma_{d}$ and $\Omega_d$ are the surface density and orbital frequency at $R_d$. 
This equation has been derived by \cite{Terquem1998}, \cite{Bate2002}, and \cite{Lai2014}. 
The rigid body approximation is a good approximation as long as the disc can communicate through wave \citep{Larwood1996}
or viscous torques \citep{WijersPringle1999}. With $\alpha_s$=1 in our simulations and the relationship 
\begin{equation}
\frac{\Omega_{d}}{\Omega_{p}}=\left(\frac{M_{*} }{M_{*}+M_{p}}\right)^{1/2}\left(\frac{R_{p}}{R_{d}}\right)^{3/2}\,,
\end{equation}
we can derive
\begin{equation}
\frac{\omega_{rb}}{\Omega_{p}}= \frac{3}{8} {\rm cos}(i_p)q \left(\frac{1}{1+q}\right)^{1/2}\left(\frac{R_{d}}{R_{p}}\right)^{3/2}\,.\label{eq:wrbfull}
\end{equation}
In our simulations with $q\ll 1$ where the gap inner edge is close to the planet position ($R_{d}\sim R_{p}$), we have
\begin{equation}
\frac{\omega_{rb}}{\Omega_{p}} \sim \frac{3}{8} {\rm cos}(i_p)  q \,,\label{eq:precessionrate}
\end{equation}
and the precession timescale is $T_{pre}=1/\omega_{rb}$.  When the inner disc is small and $R_d$ is significantly smaller than $R_p$, the precession rate can be a lot lower than the estimate from Equation \ref{eq:precessionrate}, since $\omega_{rb}$ in Equation \ref{eq:wrbfull} has $(R_d/R_p)^{3/2}$ dependence.

Now, let's derive under what condition the disc breaking can occur at $R_p$.
We follow \cite{LubowOgilvie2000} and adopt the complex representation for $\bf{l}$
and $\bf{G}$, which are $W=l_x+i l_y$ and $G=G_x+i G_y$. Equations \ref{eq:tilt1} and \ref{eq:tilt2} become
\begin{eqnarray}
\Sigma R^{2}\Omega\left[\frac{\partial W}{\partial t}-i\left(\frac{\Omega^2-\Omega_{z}^2}{2\Omega}\right)W\right]=\frac{1}{R}\frac{\partial G}{\partial R}\,\label{eq:tiltcomplex}\\
\frac{\partial  G}{\partial t}-i\left(\frac{\Omega^2-\kappa^2}{2\Omega}\right) G +\alpha \Omega G=\frac{PR^3\Omega}{4}\frac{\partial W}{\partial R}\label{eq:tilt2complex}\,,
\end{eqnarray}
where $\Omega$ and $\kappa$ are defined as Equation (25) and (26) in \cite{LubowOgilvie2000}, and $\Omega_z$ is
\begin{equation}
    \Omega_z^2=\frac{GM_*}{R^3}-\frac{GM_p}{2R_p^2 R}b_{3/2}^{(1)}\left(\frac{R}{R_p}\right)\,,
\end{equation}
where $b_{3/2}^{(1)}$ is the Laplace coefficient as in \cite{LubowOgilvie2000}.
We note that both $\frac{\Omega^2-\Omega_{z}^2}{2\Omega}$ and $\frac{\Omega^2-\kappa^2}{2\Omega}$ are on the order of $\omega_{rb}$ when the axisymmetric contribution of the companion potential has been taken into account.

When the disc breaking does not occur, the whole disc including both the inner and outer discs precesses as a whole and the resulting precession rate $\omega_p$ is much smaller
than $\omega_{rb}$ in Equation \ref{eq:precessionrate}. Thus, we drop the time dependent terms and integrate Equation \ref{eq:tiltcomplex} from $0$ to $R$:
\begin{equation}
   G=  -i\int_{0}^{R} \Sigma R'^{3}\left(\frac{\Omega^2-\Omega_{z}^2}{2}\right)W dR'\,.
\end{equation}
If we plug this $G$ into Equation \ref{eq:tilt2complex} and also drop the time dependent terms, we have
\begin{equation}
    \frac{\partial W}{\partial R}=-\frac{4}{PR^3\Omega}\left(\frac{\Omega^2-\kappa^2}{2\Omega}  + i \alpha \Omega \right) \int_0^{R} \Sigma R'^{3}\left(\frac{\Omega^2-\Omega_{z}^2}{2}\right)W dR' \,.
\end{equation}
If we replace the $\Omega^2$ terms with $\omega_{rb}$ and use some averaged surface density within $R$ (labled as $\overline{\Sigma}$) to simplify the integral from 0 to R, we have
\begin{equation}
    \frac{\partial W}{\partial R}\sim-\frac{4}{P}\left(\omega_{rb}  + i \alpha \Omega \right) \overline{\Sigma} R \omega_{rb} W \,.
\end{equation}
In order for the inner and outer discs to precess together as one ridge body, $|W|$ needs to be a constant or $|\partial W/\partial R\times R|$ needs to be small throughout the disc. In other words, $|\partial W/\partial R\times R|<W$ for all $R$.
For orders of magnitude argument, we thus have the rigid body precession condition as
\begin{equation}
    \frac{4 \overline{\Sigma} R^2}{P}\left|\left(\omega_{rb}\right)^2  + i \alpha \Omega \omega_{rb}\right| <1 \quad\quad {\rm for}\quad {\rm all}\quad R\,.
\end{equation}
This equation is most difficult to be satisfied at the radius where $P$ is small.
In our planet-disc interaction problem, $P$ is the smallest at the deepest part of the gap (at $\sim R_p$) where $P=c_{s}^2\Sigma_{gap}$.
The rigid rotation condition is then
\begin{equation}
    \frac{4 R_p^2 \overline{\Sigma}}{c_{s}^2\Sigma_{gap}}\left|\left(\omega_{rb}\right)^2  + i \alpha \Omega \omega_{rb}\right| <1 .
\end{equation}
Our simulations later will show that the disc cannot break in a highly viscous disc since the gap is shallow. Thus
we will limit ourselves to inviscid discs where the breaking condition is then 
\begin{equation}
     \frac{\Sigma_{gap}}{ \overline{\Sigma}} < \frac{4 R_p^2}{c_{s}^2}\left(\omega_{rb}\right)^2\,,\label{eq:breaking}
\end{equation}
or
\begin{equation}
     \frac{\Sigma_{gap}}{ \overline{\Sigma}} < \frac{4 \left(\frac{\omega_{rb}}{\Omega_p}\right)^2}{h^2}\,.\label{eq:breaking2}
\end{equation}
Since $\overline{\Sigma}(R_p)$ represents the averaged disc surface density at the inner disc within $R_p$, $\overline{\Sigma}(R_p)$ is $\sim\Sigma_0(R_p)$ so that $\Sigma_{gap}/\overline{\Sigma}$ can be considered as the gap depth.

If the disc is smooth without a gap so that $\Sigma_{gap}\sim\Sigma_0$, Equation \ref{eq:breaking} becomes the traditional disc breaking condition that the sound crossing time is longer than the precession timescale \citep{Larwood1996}.
When a deep gap is formed, Equation \ref{eq:breaking} can also be written as
\begin{equation}
     \frac{2 R}{c_{s}}>\sqrt{\frac{\Sigma_{gap}}{ \overline{\Sigma}}} \frac{1}{\omega_{rb}} \,,\label{eq:breaking3}
\end{equation}
which suggests that the breaking condition is that the sound crossing time is longer than the product of the precession timescale and the square root of the gap depth. The surface density factor reflects that less surface density communicates different parts of the disc less efficiently.

In this section, we  derive the gap depth and width for misaligned planets (Equations \ref{eq:depthwidth} and \ref{eq:newK}), the migration and inclination damping
rates for planets in gaps (Equations \ref{eq:ptr} to \ref{eq:tincgap}), and the disc breaking condition (Equation \ref{eq:breaking} and \ref{eq:breaking2}). In the next two sections, we will carry out numerical
simulations to test these equations.

\section{Methods}

\begin{figure*} 
\centering
\includegraphics[trim=0cm 3cm 0cm 0cm, width=0.9\textwidth]{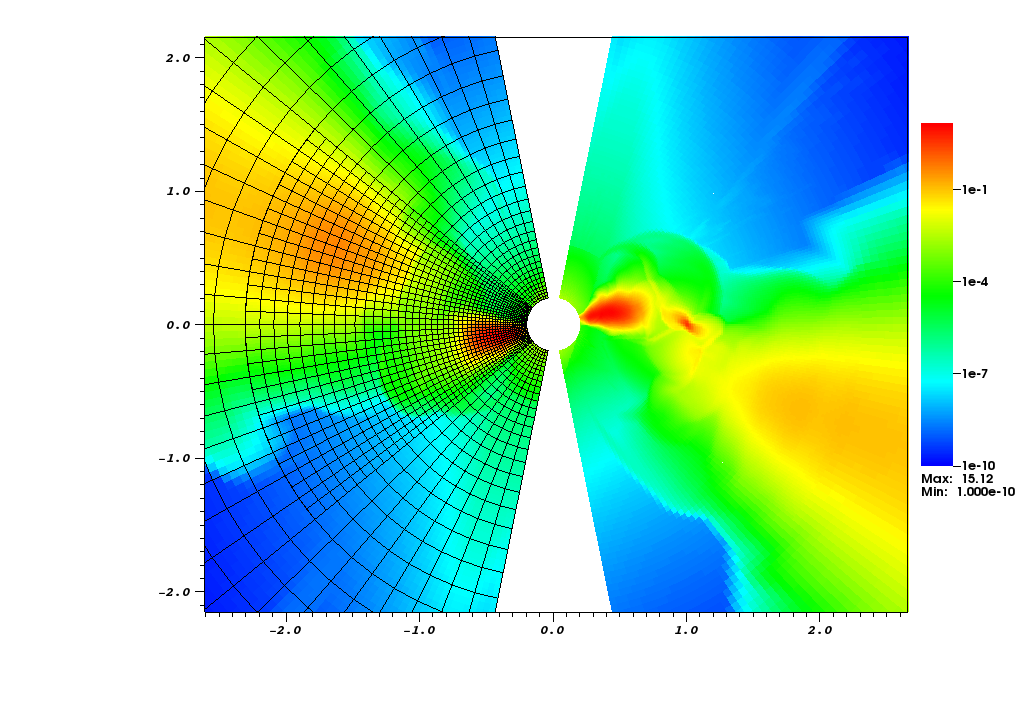} 
\caption{The density map for the  \texttt{P10I19A0} case at t=180 $T_{p}$. The left half of the plot also includes the mesh structure, with
mesh refinement close to the disc midplane. The planet and the circumplanetary disc can be seen at x=1, y=0. Initially, the inner disc
aligns with the outer disc with a common inclination of 19$^o$. But after 180 $T_{p}$ the inner disc precesses $\sim$ 180$^o$  and the misalignment between 
the inner and outer discs is now $\sim$ 38$^o$. }
\vspace{-0.1 cm} \label{fig:visit0002}
\end{figure*}

We solve the compressible Navier-Stokes equations using Athena++ (Stone et al. 2018, in preparation). 
Athena++ is a newly developed grid based magnetohydrodynamic code using a higher-order Godunov scheme for MHD and 
the constrained transport (CT) for magnetic fields. 
Compared with its predecessor Athena \citep{GardinerStone2005,GardinerStone2008,Stone2008},  Athena++ is highly optimized for 
speed and uses a flexible grid structure that enables mesh refinement, allowing global
numerical simulations spanning a large spatial range. 
Furthermore, the geometric source terms in curvilinear coordinates 
(e.g. in cylindrical and spherical-polar coordinates) are carefully implemented so that angular momentum
is conserved to machine precision, an important feature for studying the disc precession.

\begin{figure*} 
\centering
\includegraphics[trim=0cm 7cm 0cm 0cm, width=0.8\textwidth]{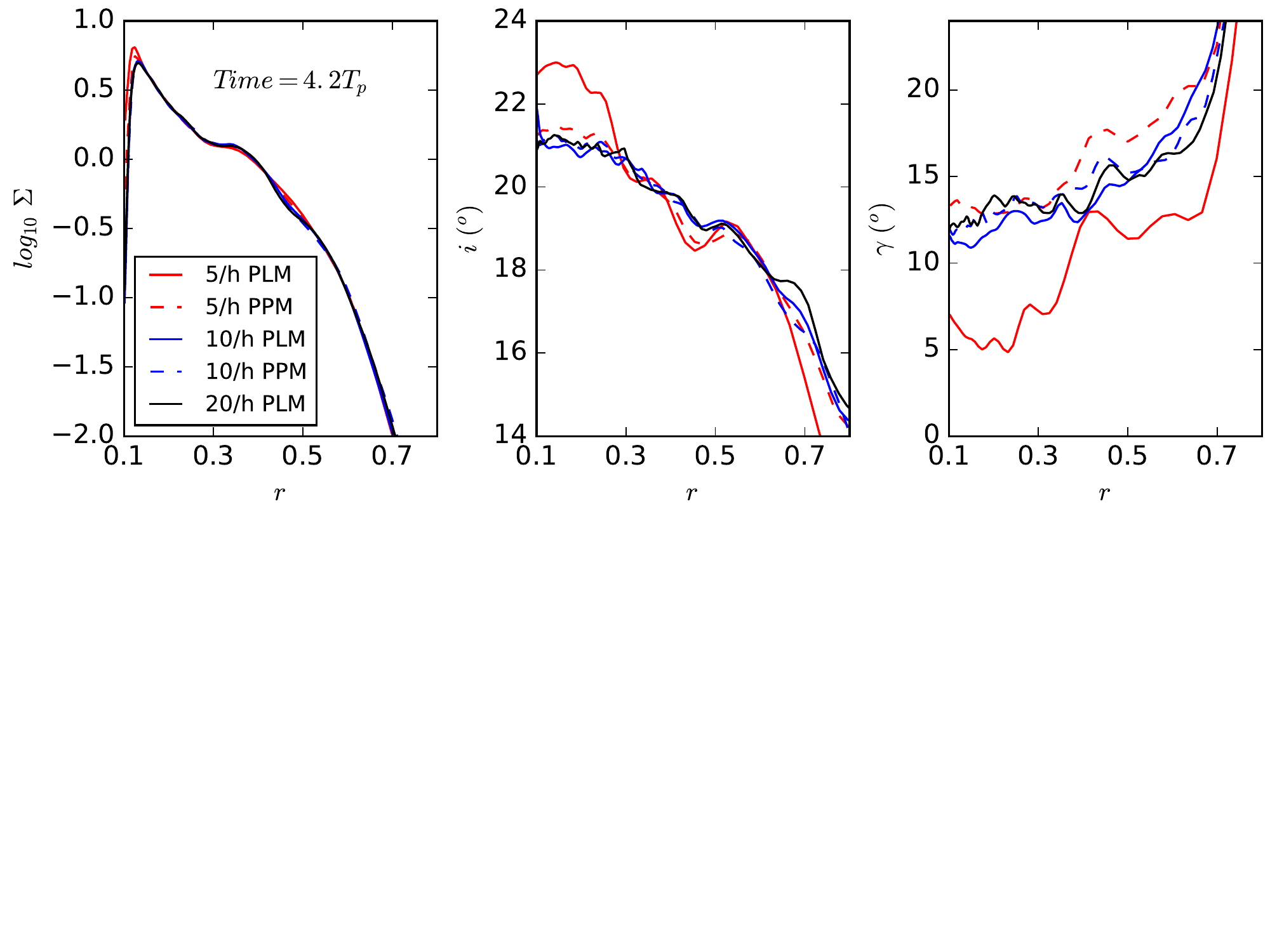} 
\caption{The resolution test with a $q=0.03$ planet on a 19$^o$ inclined orbit.  From left to right, the profiles of
the disc surface density, tilt angle, and twist angle are presented. Simulations with different resolutions and
reconstruction schemes (PLM or PPM) are presented. The simulation with both 5 grids per scale height and the PLM reconstruction scheme
behaves differently from other runs, indicating its insufficient resolution. }
\vspace{-0.1 cm} \label{fig:inclinationcompare}
\end{figure*} 

We adopt the spherical-polar coordinate system ($r$, $\theta$, $\phi$) in our simulations, with $r$ from 0.2 to 10,  $\theta$ from 0.2 to $\pi$-0.2, and 
$\phi$ from 0 to $2\pi$. 
In the radial direction, we have 160 grid points uniformly in log($r$). In the $\theta$ and $\phi$ directions,
we have 132 and 256 uniform grid points. One level of mesh refinement is used towards the disc midplane, allowing us to 
accurately simulate the disc evolution but also not to be limited by the small timestep at the disc atmosphere. 
The grid structure is shown in Figure \ref{fig:visit0002}.
The outflow boundary condition\footnote{This outflow condition is different from the default outflow condition in the code.  We follow the standard practice for the outflow boundary condition \citep{StoneNorman1992}: if $v_{r}$ in the ghost zones are pointing to the active zones, they are set to be zero to limit the inflow.}, reflecting boundary condition, and periodic boundary condition
are adopted in the $r$, $\theta$, and $\phi$ direction respectively.

Although we adopt spherical-polar coordinates for the simulations, we 
use cylindrical coordinates to set up the initial condition.
In this paper, we use ($R$, $\phi$, $z$) to denote positions in cylindrical coordinates and
($r$, $\theta$, $\phi$) to denote positions in spherical-polar coordinates.  In both coordinate systems, 
$\phi$ always represents the azimuthal direction (the direction of disc rotation). 

The initial density profile at the disc midplane is
\begin{equation}
\rho_{0}(R,z=0)=\rho_{0}(R_{0},z=0)\left(\frac{R}{R_{0}}\right)^p\,,
\end{equation}
where $p$ is set to be -2.25.
$R_{0}$ equals the planet's semimajor axis ($R_p$) and $R_0=R_p=1$ in code units.
The temperature is assumed to be constant on cylinders:
\begin{equation}
T(R,z)=T(R_{0})\left(\frac{R}{R_{0}}\right)^q\,,\label{eq:temperature}
\end{equation}
where $q$ is set to be -0.5. With our choice of $p$ and $q$, the disc surface density in our simulations follows $\Sigma\propto R^{-1}$. 
For our main set of simulations, $T(R_0)$ is chosen so that the disc scale height $H=c_{s}/\Omega$ at $R_0$ is 0.1, where
$c_{s}=\sqrt{p/\rho}$ is the isothermal sound speed.
This disc scale height
at $R_0$ is resolved by 10 grid points with out default resolution. 

Hydrostatic equilibrium in the $R-z$ plane requires that (e.g. \citealt{nelson2013})
\begin{equation}
\rho_{0}(R,z)=\rho_{0}(R,z=0) {\rm exp}\left[\frac{GM}{c_{s}^2}\left(\frac{1}{\sqrt{R^2+z^2}}-\frac{1}{R}\right)\right]\,,\label{eq:rho0}
\end{equation}
and
\begin{equation}
v_{\phi}(R,z)=v_{K}\left[(p+q)\left(\frac{c_{s}}{v_{K}}\right)^2+1+q-\frac{qR}{\sqrt{R^2+z^2}}\right]^{1/2}\,,\label{eq:vphi}
\end{equation}
where  $v_{K}=\sqrt{GM_{*}/R}$. 
For thermodynamics, we adopt the adiabatic equation of state with $\gamma$=1.4, but with a short cooling time as in \cite{Zhu2015c}. We have chosen the cooling time as 0.01 of the local orbital time, which is motivated by the realistic cooling time for disc regions at $\sim$100 AU \citep{Zhu2015c}. Discs with such short cooling time behave very similarly to the disc with the locally isothermal equation of state. 

To simulate the interaction between the misaligned planet and the disc, we can either (1) set the disc midplane at the midplane of the grid (the $\theta=\pi/2$ plane) and put the planet on an inclined orbit, as done in \cite{Arzamasskiy2018}, or (2) we can set the planet's orbital plane at the midplane of the grid and tilt the disc with respect to the grid midplane. The setup (1) ensures that the disc's Keplerian motion is along one of the grid directions ($\phi$ direction) so that it minimizes the grid noise for simulating a Keplerian disc. Thus, we adopted this setup in  \cite{Arzamasskiy2018}, where the low mass planet does not affect the disc evolution much. However, in current work with a high mass planet, the inner disc can undergo significant nodal precession around the total angular momentum vector of the whole system. Since the disc's mass is assumed to be negligible compared with the planet's mass in our simulations, the total angular momentum vector equals the planet's orbital angular momentum vector. Then, with the setup (1), the disc can move away from the grid midplane during the nodal precession. For example, if the planet is
45 degrees inclined to the disc plane, the disc will move away from the midplane to the polar region of the grid after it precesses for 180$^o$. Although the grid midplane has small numerical diffusion, the polar region
has the lowest resolution and largest numerical diffusion. 
 Considering this disadvantage, in this work we adopt the setup (2) where the planet's orbital plane is at the grid midplane and the disc is tilted with respect to the grid midplane (Figure \ref{fig:visit0002}). In this setup,
the total angular momentum vector is along the polar direction, and the disc's angular momentum vector keeps the same angle from the gird polar direction during the disc's precession so that the tilt between the disc and the grid midplane remains a constant and the numerical error won't change dramatically with time.

Suppose that the disc is tilted by an angle of $i_d$, the coordinates of any point in the coordinate system that has the tilted disc as the midplane ($r'$, $\theta'$, $\phi'$) are related to its coordinates in the numerical grid ($r$, $\theta$, $\phi$):
\begin{eqnarray}
    \sin^2 \theta'=\cos^2 \phi \cdot \sin^2\theta+\cos^2 (i_d) \cdot \sin^2\phi \cdot \sin^2\theta\nonumber\\
    +\sin^2 (i_d) \cdot \cos^2\theta-2 \cos (i_d) \cdot \sin (i_d) \cdot \cos\theta \cdot \sin\theta \cdot \sin\phi\,.
\end{eqnarray}
where $i_d$ is the disc inclination. When $i_d>0$, the disc's angular momentum vector is pointing to the positive $y$ direction in the numerical grid. The Cartesian coordinates ($x$, $y$, $z$) correspond to the spherical coordinates ($r$, $\theta$, $\phi$), while   ($x'$, $y'$, $z'$) correspond to  ($r'$, $\theta'$, $\phi'$).
With $\theta'$ calculated for each grid point at ($r$, $\theta$, $\phi$), we can calculate $R'$ and $z'$ with $R'=r \sin \theta'$ and $z'=r \cos \theta'$, and then use Equations \ref{eq:temperature}, \ref{eq:rho0},  and \ref{eq:vphi} to calculate the temperature, density, and velocities for every grid point ($R$ and $z$ in these equations are now $R'$ and $z'$).

Our main suite of simulations include 3$\times$3$\times$2=18 simulations with 3 different disc inclinations ($i_d$=-0.17, -0.34, -0.68 \footnote{So the disc's angular momentum vector is pointing to the negative $y$ direction in the initial direction}, or equivalent to the planet inclination $i_p=$0.17, 0.34, 0.68 or 10$^o$, 19$^o$, 39$^o$ ), 3 viscosity parameters ($\alpha$=10$^{-3}$, 10$^{-2}$, and 10$^{-1}$), and 2 planet masses ($q=0.003$ and 0.01, or 3 $M_{J}$ and 10 $M_J$). The planet is fixed to be on a circular orbit and its potential has a smoothing length of 0.6 $H$. As shown by \cite{Rein2012}, dynamical friction has a weak dependence on the smoothing length. We label each simulation in the following manner: \texttt{P10I10AM1} means a 10 Jupiter mass planet (or $q=0.01$)
on a 10$^o$ inclined orbit in an $\alpha=10^{-1}$ disc (\texttt{M1} means minus 1). 

We  also carry out two additional simulations to complement the main suite of simulations: 1) one inviscid simulation ($\alpha=0$) with $i_p$=19$^o$ and $M_p=10 M_J$ (\texttt{P10I19A0}) to be compared with the corresponding viscous simulations, 2)  a thin disc with $h$=0.05, $\alpha=10^{-3}$, $M_p$=10 $M_{J}$, and  $i_p$=39$^o$(\texttt{THINP10I39AM3}) to study disc breaking in a thin disc.

\subsection{Numerical Convergence}
To test the numerical convergence of our setup, we also carry out a suite of inviscid simulations using different resolutions and numerical reconstruction schemes. These simulations have the same setup as our fiducial cases with $i_p$=19$^o$, except that the planet mass is $q=0.03$ and the radial extend is from $r_{min}=0.1$ to $r_{max}=1$. The larger planet mass makes the disc undergo significant precession during a shorter period of time. 
The resulting radial profiles of
the disc surface density ($\Sigma$), tilt ($i$), and twist ($\gamma$) at 4.2 planetary orbits (4.2 $T_p$) are presented in Figure \ref{fig:inclinationcompare}. 
Different resolutions and
reconstruction methods (second-order piecewise linear method: PLM,  or third-order  piecewise parabolic method: PPM) are explored.

The tilt vector at each radius (${\bf l}(R,t)$ in Equation \ref{eq:tilt1}) characterizes how a disc is warped. 
The tilt angle ($i$) and twist angle ($\gamma$) at radius $R$ can be calculated using $\bf{l}$:
\begin{eqnarray}
    i=\cos^{-1}(l_{z})\\
    \gamma=\cos^{-1}\left(\frac{-l_y}{l_x^2+l_y^2}\right)\,.
\end{eqnarray}
Our definition of twist is to ensure that the initial twist angle is zero. We also define that the twist angle increases
in the clockwise direction in the $x-y$ plane. 

Figure \ref{fig:inclinationcompare} shows that the simulation with both the PLM reconstruction scheme and 5 grids per scale height at $R_{p}$ has very different $i$ and $\gamma$
profiles compared with simulations having a higher resolution or the PPM reconstruction scheme. Thus, we conclude that, if we use the PLM reconstruction
scheme, we need
at least 10 grids per scale height. If we use the PPM reconstruction scheme, 5 grids per scale height is sufficient.  On the other hand,
we notice that the simulation having both 5 grids per scale height and the PPM scheme 
still has some minor differences from the higher resolution runs. Thus,
we decide to use 10 grids per scale height with the PLM scheme as our fiducial setup. On the other hand, with our fiducial setup, the resolution is only 5 grids per scale height for the thin disc simulation \texttt{THINP10I39AM3}.  Thus, we use the PPM scheme for this thin disc simulation.

\section{Results}

\begin{figure} 
\centering
\includegraphics[trim=0cm 1cm 0cm 0cm, width=0.5 \textwidth]{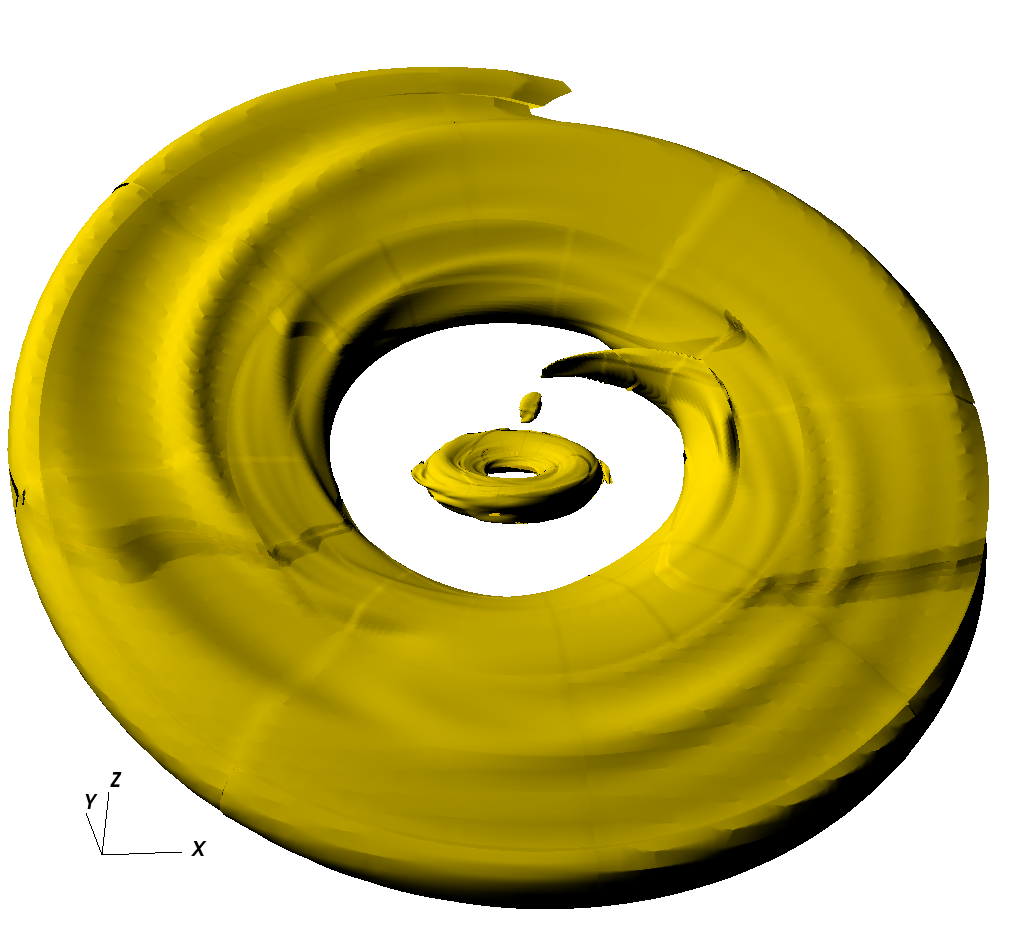} 
\caption{ The same as Figure \ref{fig:visit0002} but shows the isodensity surface. The tilted circumplanetary disc is also shown. 
The movie can be downloaded from \url{http://www.physics.unlv.edu/~zhzhu/Media/inclinedplanets.mpg}}
\vspace{-0.1 cm} \label{fig:visit0003}
\end{figure}

\begin{figure*} 
\centering
\includegraphics[trim=0cm 0cm 0cm 0cm, width=0.8\textwidth]{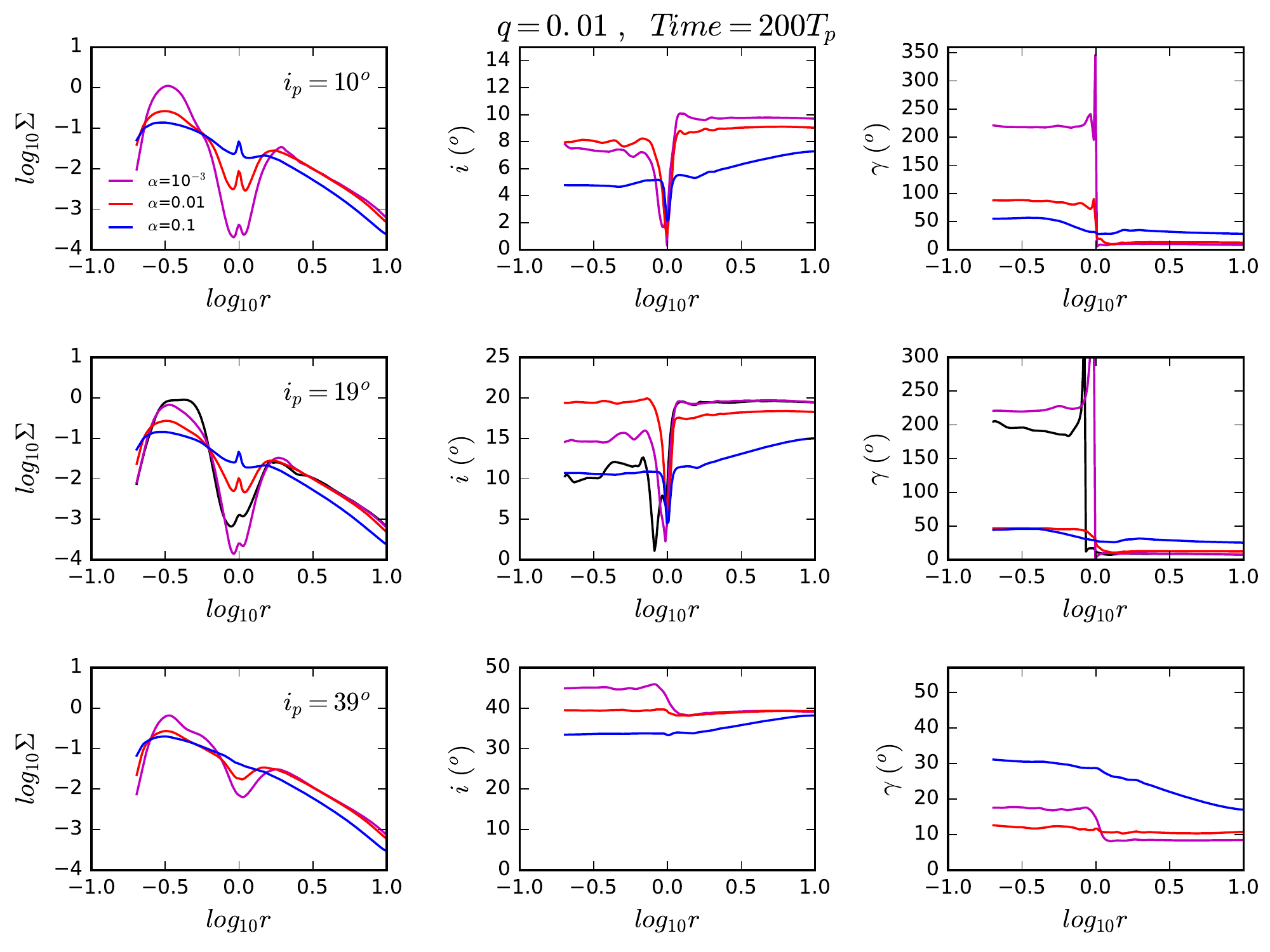} 
\caption{ The  profiles of  the azimuthally averaged surface density (left panels), disc tilt (middle panels), and twist (right panels) 
for cases with a q=0.01 planet in the disc at the  time of 200 $T_{p}$. The planet is on a $10^o$, $19^o$, and $39^o$ inclined orbit from top rows to bottom rows. 
The purple, red, and blue curves represent discs with $\alpha=10^{-3}$, $10^{-2}$,
and $10^{-1}$. The black curves in the $i_{p}=19^o$ panels are from the inviscid simulation with zero viscosity. }
\vspace{-0.1 cm} \label{fig:inclinationcompare10MJ}
\end{figure*}

\begin{figure*} 
\centering
\includegraphics[trim=0cm 0cm 0cm 0cm, width=0.8\textwidth]{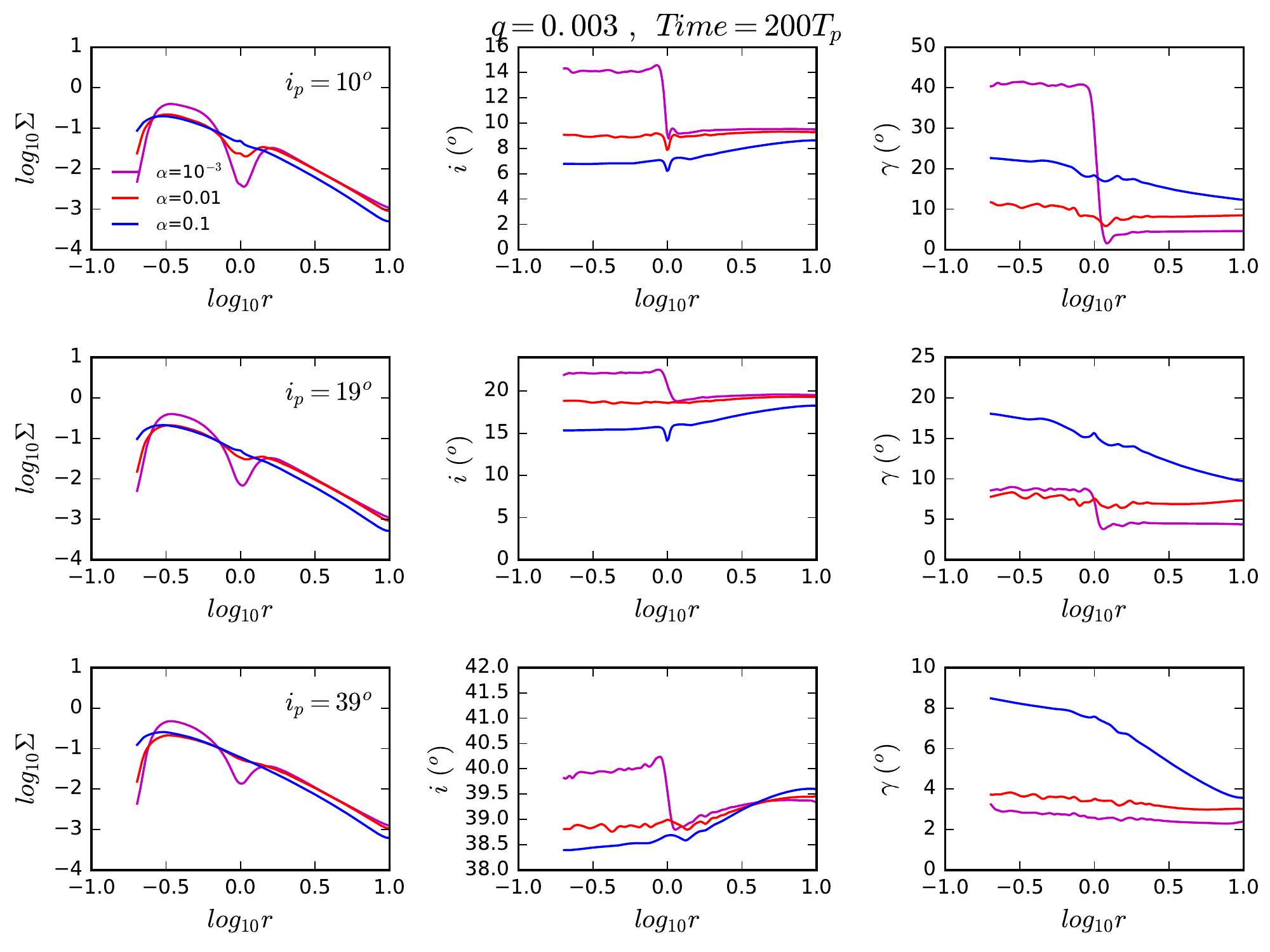} 
\caption{ Similar to Figure \ref{fig:inclinationcompare10MJ} but for a $q=0.003$ planet in the disc.}
\vspace{-0.1 cm} \label{fig:inclinationcompare3MJ}
\end{figure*}

We run the simulations for 200-300 planetary orbits. The poloidal cut for the disc density structure of  \texttt{P10I19A0} is shown 
in Figure \ref{fig:visit0002}. The planet, together with the circumplanetary region, can be seen on the right at the grid midplane. Initially, both the inner and outer discs are  tilted by 19$^o$. However, the massive planet has carved out a very deep gap   so that the inner and outer discs break from each other and precess at different precession rates.
After 180 planetary orbits, the inner disc precesses for $\sim$180$^o$ around the z-axis and it is now
facing the other direction. Thus, the angle between the angular momentum vectors of the inner and outer discs  is now 19$^o \times$ 2=38$^o$.
The isodensity surface for this snapshot is shown in Figure \ref{fig:visit0003}. We can clearly see that the inner and outer discs are misaligned. Spirals are apparent at both the inner and outer discs. Another interesting phenomenon in both Figure \ref{fig:visit0002} and \ref{fig:visit0003} is that the circumplanetary disc (CPD) around the planet is aligned with the outer disc 
instead of the inner disc. This may be because the
CPD forms by accreting material from the outer disc and this incoming material carries the angular momentum of the outer disc. 

\subsection{Disc Structure}

\begin{figure} 
\centering
\includegraphics[trim=0cm 1cm 5.5cm 0cm, width=0.5\textwidth]{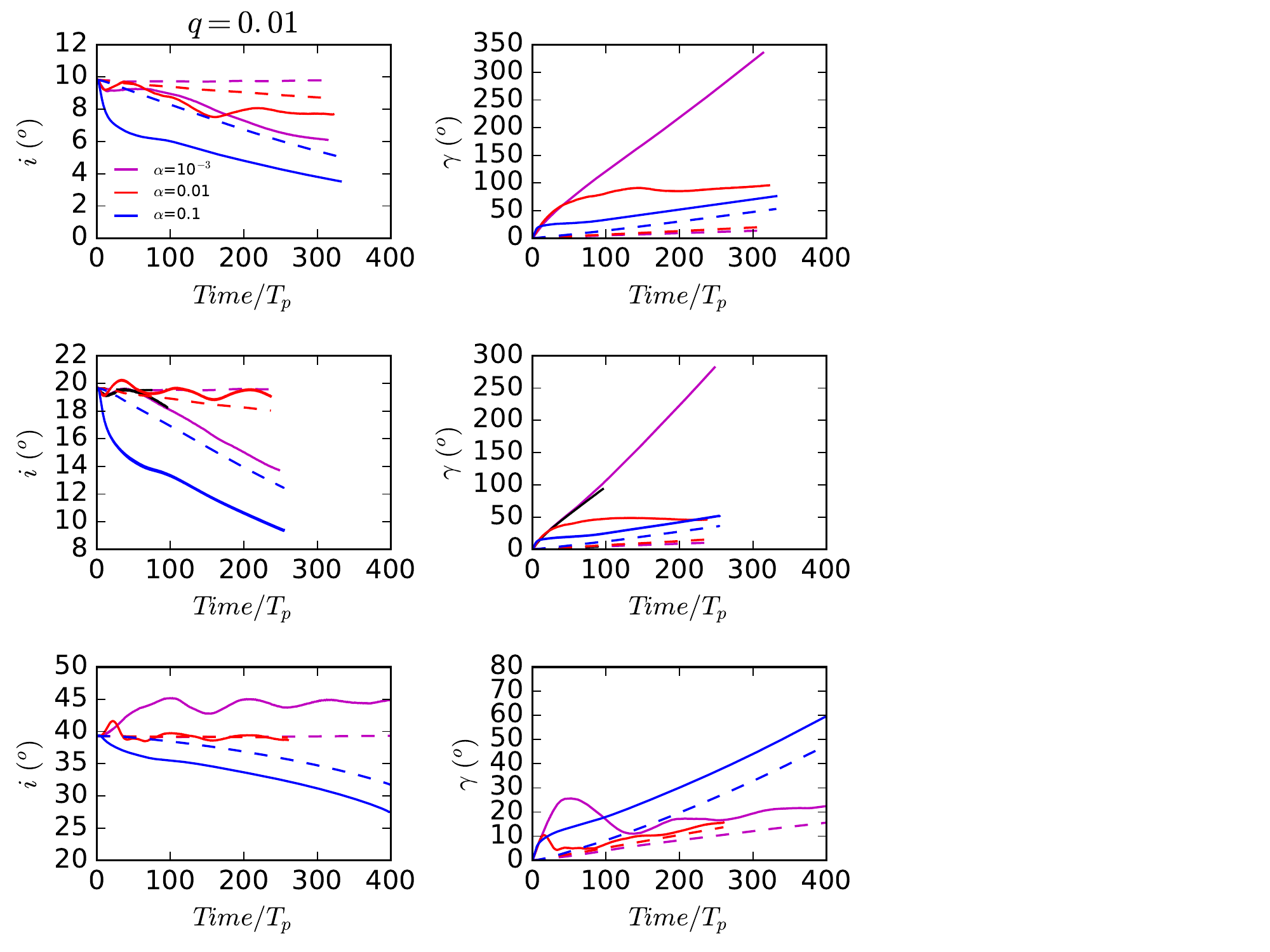} 
\caption{The time evolution of the disc tilt angle (left panels) and twist angle (right panels) when a $q$=0.01 planet is in the disc. 
The solid curves represent the inner disc within $R$=0.7, while the dashed curves represent
the outer disc beyond $R$=1.5.
The planet is on a $10^o$, $19^o$, and $39^o$ inclined orbit from top to bottom rows. 
The purple, red, and blue curves represent discs with $\alpha=10^{-3}$, $10^{-2}$,
and $10^{-1}$. }
\vspace{-0.1 cm} \label{fig:inclinationtime10MJ}
\end{figure}

\begin{figure} 
\centering
\includegraphics[trim=0cm 1cm 5.5cm 0cm, width=0.5\textwidth]{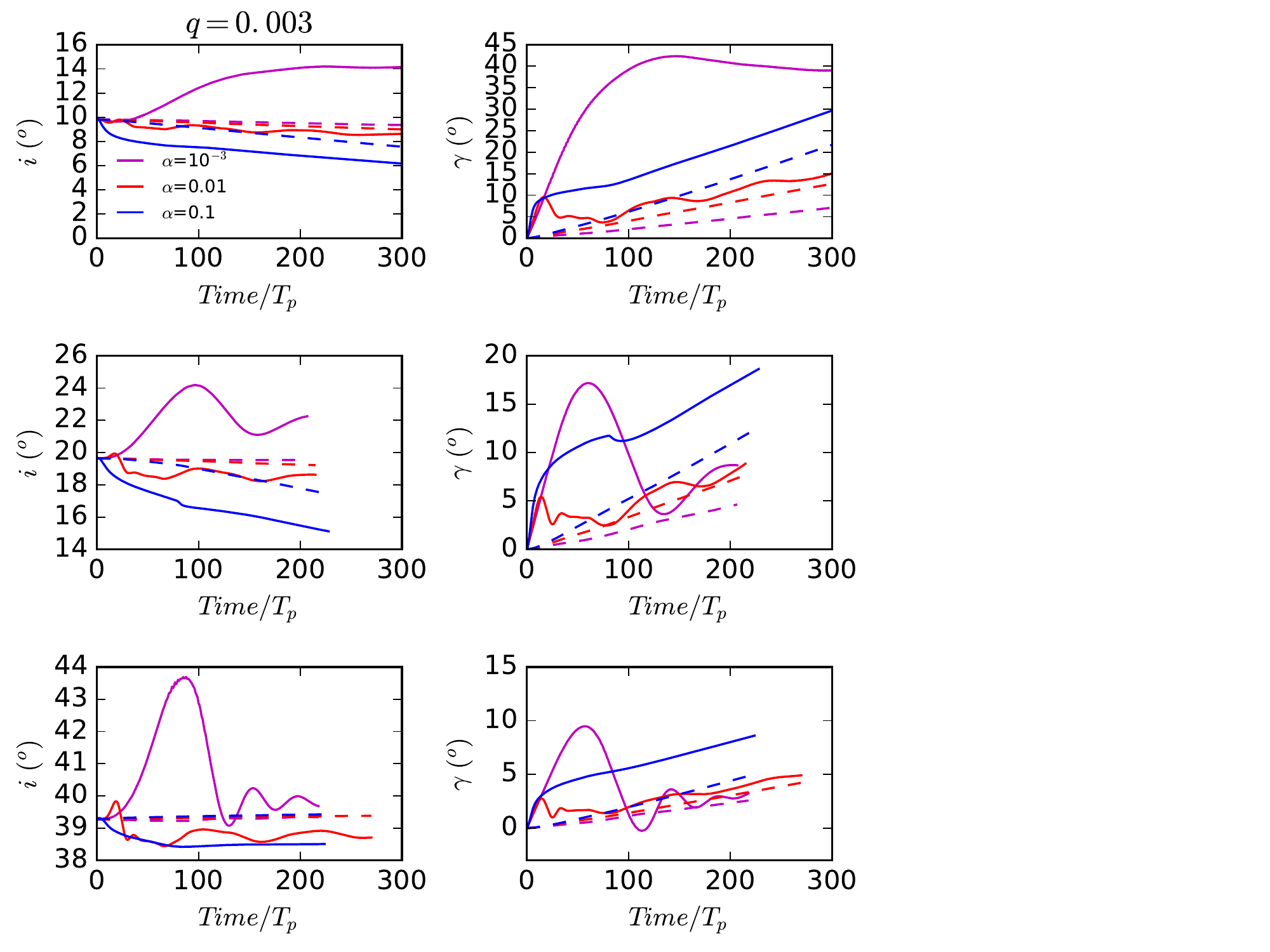} 
\caption{ Similar to Figure \ref{fig:inclinationtime10MJ} but for $q$=0.003 cases.}
\vspace{-0.1 cm} \label{fig:inclinationtime3MJ}
\end{figure}

At 200 planetary orbits,
the radial profiles of the disc surface density, tilt, and twist are shown in Figures \ref{fig:inclinationcompare10MJ}
and \ref{fig:inclinationcompare3MJ} with a 3 $M_J$ and 10 $M_J$ planet in the disc respectively. There are several noticeable trends from these figures:

1) The larger $\alpha$ the disc has, the shallower gap the planet induces. This is similar to
gap opening by coplanar planets \citep{Fung2014, Kanagawa2015}. While the planet tries to open a gap by depositing
angular momentum into the disc, the disc tries to close the gap due to viscous diffusion. Thus, a higher viscosity
leads to a shallower gap. One exception to notice  in Figure  \ref{fig:inclinationcompare10MJ} is that the inviscid case \texttt{P10I19A0}
has a shallower gap than the viscous case \texttt{P10I19AM3}. This is observed in coplanar simulations too (e.g. Zhang et al. 2018). Two effects play roles here: a)
a very low viscosity in disks can trigger instabilities (e.g. Rossby Wave Instability, \citealt{Lovelace2009}) at the gap edge, which leads to turbulence that will close the gap; b) the gap edge can also become eccentric when the viscosity is low \citep{lubow1991a,lubow1991b, kley06, Teyssandier2017}, in which case azimuthally averaging over an elliptical gap can smear out the density profile of the gap.

2) The more inclined the planet is, the shallower gap the planet induces. 
This is consistent with our derivation (Equations \ref{eq:qinclined} and \ref{eq:newK}) that the inclined planet has less gravitational 
interaction with the disc. We will compare the gap depth with Equation \ref{eq:newK} in more detail in \S 4.3. 

3) The tilt ($i$) of both the inner and outer discs 
does not changed significantly at 200 orbits. 
When $\alpha$ is large (e.g. $\alpha$=0.1),
the tilt seems to
decrease slightly towards the inner disc. When $\alpha$ is small (e.g. $\alpha$=$10^{-3}$), the tilt seems to
increase slightly towards the inner disc if the gap  is not deep (e.g. Figure \ref{fig:inclinationcompare3MJ}).

4) The deeper the gap is, the larger the difference between the twist of the inner and outer discs is.
For the three cases which have the deepest gaps ( \texttt{P10I10AM3},  \texttt{P10I19AM3},  \texttt{P10I19A0}), 
the twist difference between the inner and outer discs reaches more than 200$^o$. 
For discs with shallow gaps, the twist difference between the inner and outer discs is limited to a
small value. 

Figures \ref{fig:inclinationtime10MJ} and \ref{fig:inclinationtime3MJ} show the time evolution of
the tilt and twist of the inner (solid curves) and outer (dashed curves) discs. The inner and outer discs are defined as the
disc region smaller than $R=0.7$ and larger than $R=1.5$ respectively. Their tilt and twist angles are calculated using the angular momentum vector that has been integrated over the inner and outer discs. For the three cases that have deep gaps
( \texttt{P10I10AM3},  \texttt{P10I19AM3},  \texttt{P10I19A0}), we can see that the inner and outer discs precess at very different rates. 
The twist ($\gamma$) of the inner disc increases much faster than that of the outer disc, indicating that the two discs
break and they are not dynamically connected. The measured precession rates ($\partial \gamma/\partial t$) for the inner discs of \texttt{P10I10AM3} and  \texttt{P10I19AM3} are  $\Omega_p/340$ and $\Omega_p/300$. Considering that
the analytical prediction from Equation \ref{eq:precessionrate} is $\Omega_p/276$ with i=15$^o$ and q=0.01, the simulation results
agree well with the theory.

On the other hand, for the shallow gap cases, e.g.  \texttt{P10I19AM2},  \texttt{P3I10AM3}, the twist of the inner disc
increases fast initially. But it slows down at later times and precesses at the same rate as the outer disc. The inner and outer discs thus have a constant
relative twist angle. 
It seems that the disc tries to break but manages to maintain a twist balance between the inner and outer discs. We will discuss disc breaking more quantitatively in \S 4.3.

\subsection{Migration and Inclination Damping Rates}
Although the planet is kept on a fixed circular orbit, we can calculate the planet migration rate and the inclination damping
rate using the gravitational force exerted on the planet by the disc. Following \cite{Burns1976}, the gravitational force experienced by the planet can be decomposed
into
\begin{equation}
    {\bf F}=R \hat{e_{R}}+T \hat{e_T}+ N \hat{e_{N}}\,,
\end{equation}
where $\hat{e_R}$, $\hat{e_T}$, and $\hat{e_N}$ are unit vectors along the radial direction, planet's velocity direction, and the normal direction to the planet's orbital plan (the direction of $\hat{e_R}\times\hat{e_T}$).  
These components of the force lead to the planet's orbital evolution,
\begin{eqnarray}
\frac{d r_p}{dt}&=&2\frac{r_p^{3/2}}{(GM_*)^{1/2}m_p}T\,,\\
\frac{d i_p}{dt}&=&\frac{r_p^{1/2}}{(GM_*)^{1/2}m_p}N {\rm cos}(\Omega_p t-\gamma_{out}(t))\,,
\end{eqnarray}
where we assume that the eccentricity is zero, $i_p$ is the planet's inclination angle with respect to the outer disc, and $\Omega_p t$ represents the planet's azimuthal angle with respect to the x-axis on the planet's orbital plane. The quantity $\gamma_{out}(t)$ is the outer disc's tilt angle at time $t$, and thus it is the angle from the x-axis to the line of nodes between the planet's plane and the outer disc plane. 
 $(\Omega_p t-\gamma_{out}(t))$ is thus
the angle from the line of nodes to the planet's position vector on the planet's orbital plane. 
Note that both $r_p$ and $i_p$ are osculating elements of the orbit. Thus, they vary during one orbital time. 
To get the long-term orbital evolution of the planet, we integrate $dr_p/dt$ and $di_p/dt$ from 160 to 200 planetary orbits
to calculate the migration and inclination damping rates. 
Since the outer disc precesses slowly as shown in Figures \ref{fig:inclinationtime10MJ} and \ref{fig:inclinationtime3MJ}, we set $\gamma_{out}=0$ for these calculations. We have verified that choosing a slightly different $\gamma_{out}$ based on Figures \ref{fig:inclinationtime10MJ} and \ref{fig:inclinationtime3MJ} has negligible effect on the results.

\begin{figure} 
\centering
\includegraphics[trim=0cm 0cm 0cm 0cm, width=0.5\textwidth]{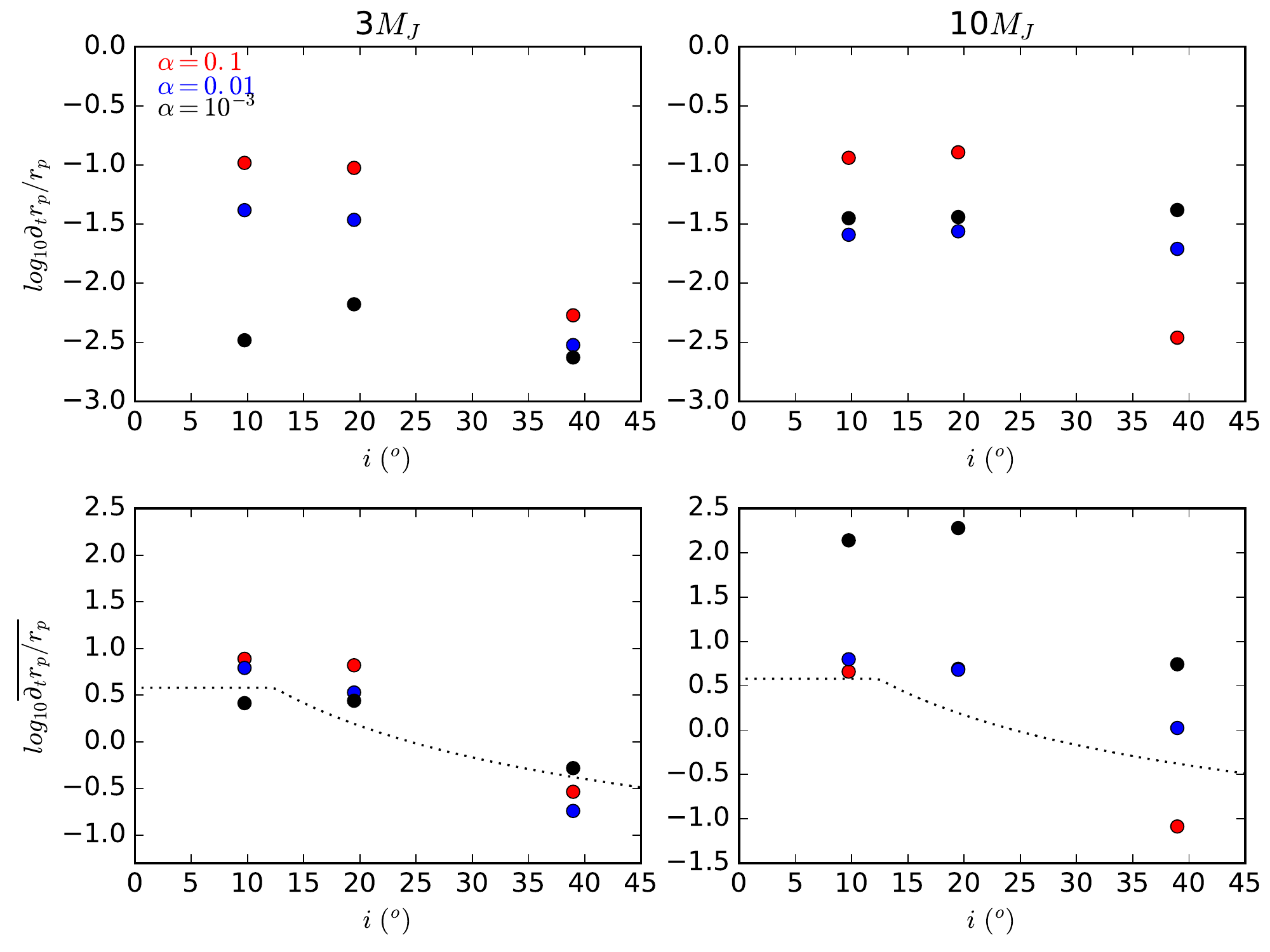} 
\caption{ The migration rates for the inclined planets with different inclination angles.  
Different colors represent the discs with different viscosity coefficients ($\alpha$).
The left two panels are for 3 $M_{J}$ mass planets ($q=0.003$), while the right two panels are for 10 $M_J$ mass planets ($q=0.01$). The bottom panels
are similar to the upper panels but the migration rates are normalized as in Equations \ref{eq:normalizedmig} and \ref{eq:normalizedinc}. The dotted curves are from the analytical theory for low mass planets.
}
\vspace{-0.1 cm} \label{fig:migration}
\end{figure}

\begin{figure} 
\centering
\includegraphics[trim=0cm 0cm 0cm 0cm, width=0.5\textwidth]{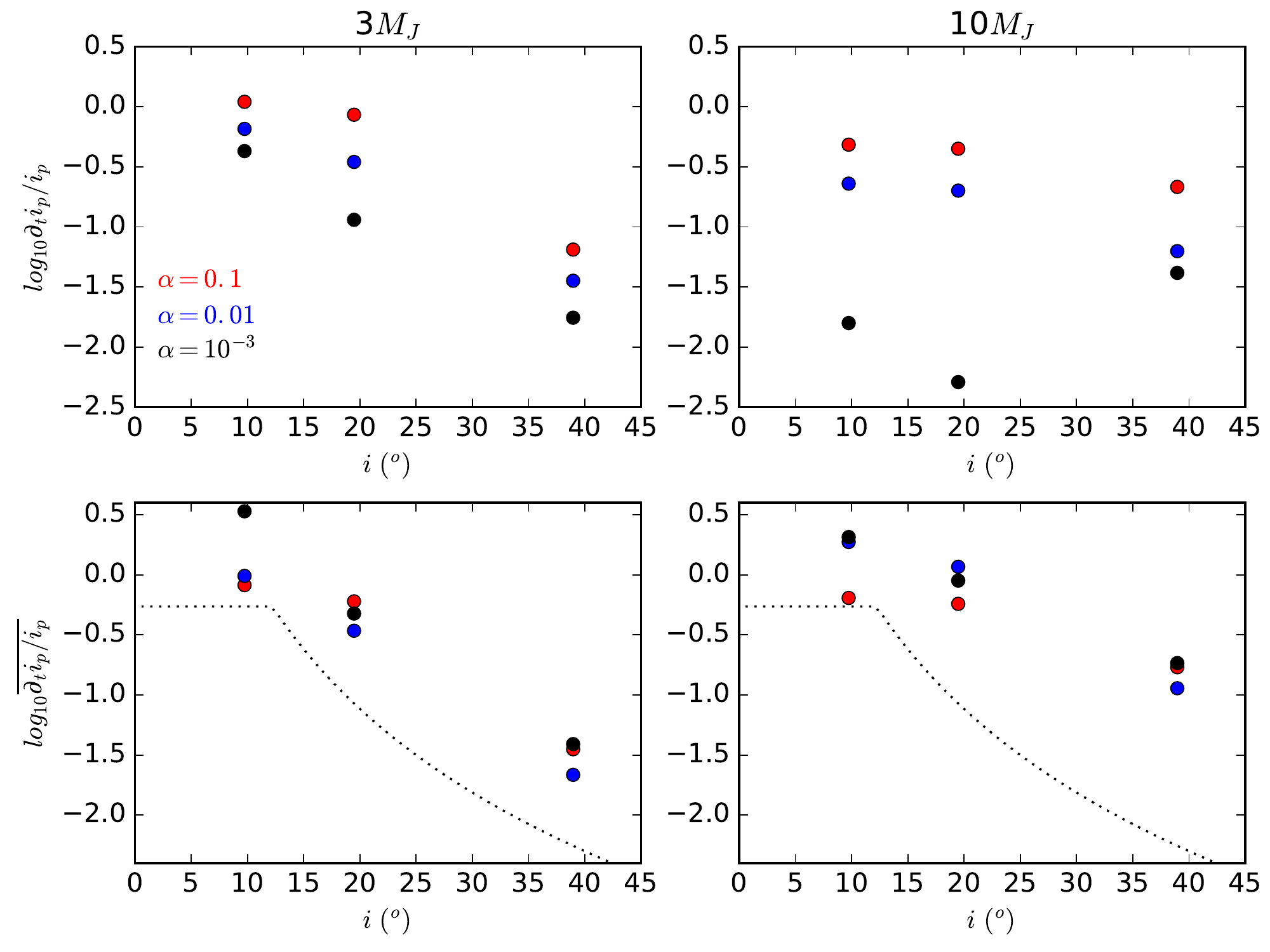} 
\caption{ Similar to Equation \ref{fig:migration} but for the inclination damping rates. }
\vspace{-0.1 cm} \label{fig:dampinc}
\end{figure}

Figures \ref{fig:migration}  and \ref{fig:dampinc} show the planets' migration and inclination damping rates in our simulations. 
Different colors represent cases with different disc viscosities. The upper panels show the measured rates in code units,
while the lower panels show the normalized rates. The normalized rates are defined as
\begin{eqnarray}
    \frac{1}{r_p}\overline{\frac{d r_p}{dt}}&=&\frac{1}{r_p}\frac{d r_p}{dt} \frac{\Sigma_0}{\Sigma_{gap}} t_{mig} \,,\label{eq:normalizedmig}\\
    \frac{1}{i_p}\overline{\frac{d i_p}{dt}}&=&\frac{1}{i_p}\frac{d i_p}{dt} \frac{\Sigma_0}{\Sigma_{gap}} t_{inc} \,,\label{eq:normalizedinc}
\end{eqnarray}
where $t_{mig}$ and $t_{inc}$ are given in Equations \ref{eq:tmig} and \ref{eq:tinc}. $\Sigma_{gap}$ is defined as
$(\Sigma(R_p-H/2)+\Sigma(R_p+H/2))/2$. We did not use the surface density at $R_p$ as the gap surface density since both the horseshoe material and the circumplanetary region lead to a density spike at $R_{p}$.
In Equations \ref{eq:normalizedmig} and \ref{eq:normalizedinc}, we normalize the rates using the gap surface density. This is because we expect the rates to be roughly proportional to the gap surface density since
the dynamical friction process, which determines the orbital evolution of a moderately inclined planet, is the gravitational interaction between the planet and the local disc where the planet is traveling through.

If a gap is not induced by a planet,
Equations \ref{eq:normalizedmig} and \ref{eq:normalizedinc} are reduced to Equations \ref{eq:ptr} and \ref{eq:ptpi} 
except that $t_{mig}$ and $t_{inc}$  are now moved to the left side. These normalized rates without gaps are plotted as the dotted curves in the bottom panels of Figures  \ref{fig:migration}  and \ref{fig:dampinc}. 

Overall, we can see that the migration and inclination damping rates decrease with a higher planet inclination 
and a deeper disc gap (lower $\alpha$).
The normalized rates agree well with the analytical formulae for most cases. But the
measured rates (especially the migration rates) are higher than the analytical formulae for 10 $M_J$ planets in $\alpha=10^{-3}$ discs where deep gaps are induced. This indicates that simply scaling
the rates with the smallest density in the gap is not adequate.
When the planet travels across the gap, its gravity not only interact with the deepest part of the gap but also with the gap edge
and even the full disc region which has much higher density. Thus, the relationships between the rates and the gap depth may not be linear.

\begin{figure} 
\centering
\includegraphics[trim=0cm 0cm 8cm 0cm, width=0.5\textwidth]{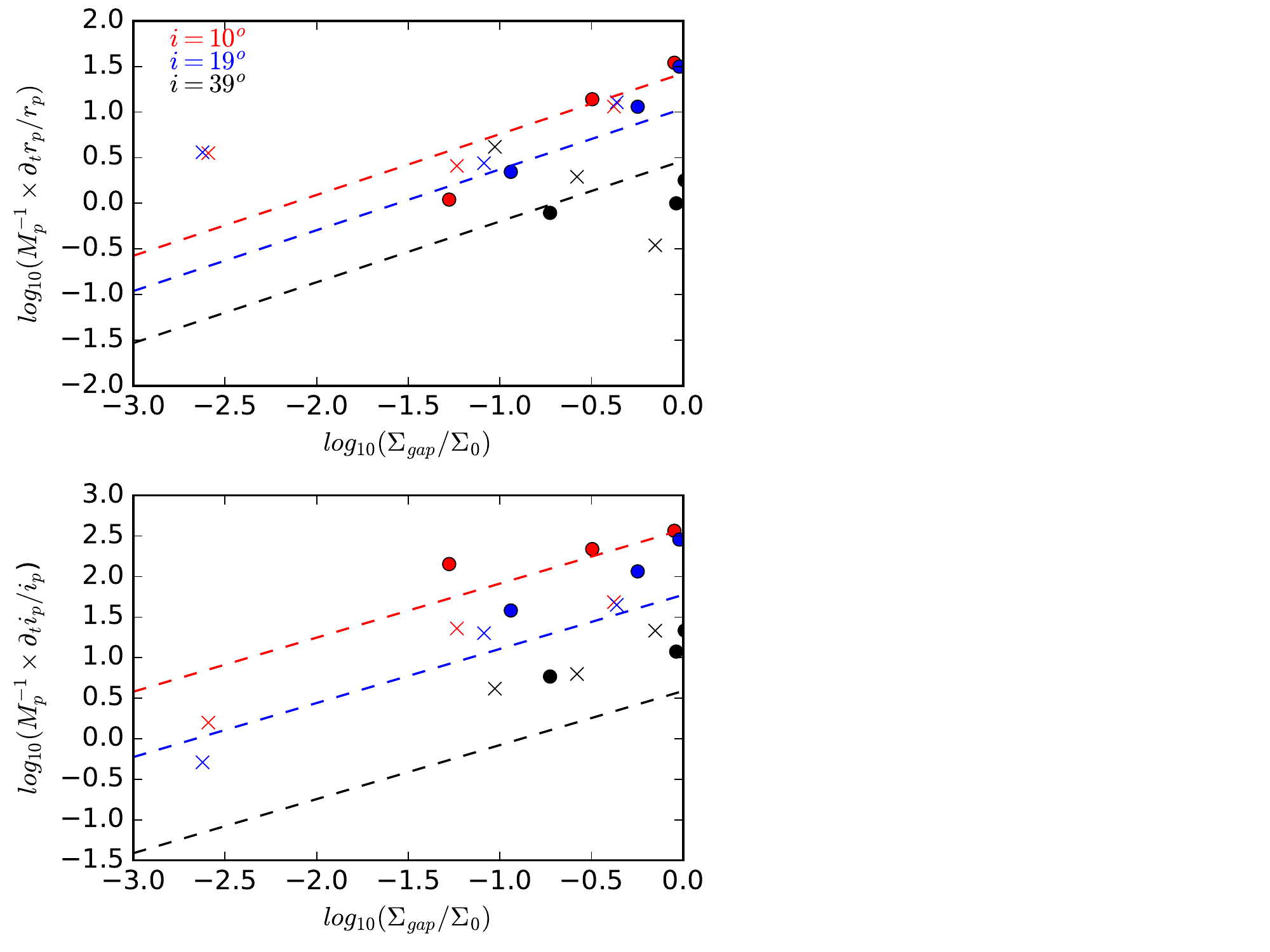} 
\caption{ The migration rate (the upper panel) and the inclination damping rate (the lower panel) with respect to the gap depth. 
Different colors represent different planet inclinations. The dots are for 3 $M_{J}$ planet cases while the
crosses are for 10 $M_{J}$ planet cases .}
\vspace{-0.1 cm} \label{fig:migincmass}
\end{figure}

On the other hand, we can use simulation results to seek empirical relationships between the gap depth and the planet's orbital evolution.
Figure \ref{fig:migincmass} summarizes the planet migration and inclination damping rates with respect to the gap depth.
The dashed lines are from the analytical formulae (Equations \ref{eq:tmiggap} and \ref{eq:tincgap}) using $qs=2/3$ and parameters in the simulations ($\Sigma_0$
is set to be 0.07 as inferred from Figures \ref{fig:inclinationcompare10MJ} and \ref{fig:inclinationcompare3MJ}).

If the rates are proportional to $\Sigma_{gap}$, the fudge factor $qs$ in Equations \ref{eq:tmiggap} and \ref{eq:tincgap} is 1. 
Based on the argument above that a misaligned planet can also interact
with the disc material beyond the center of the gap, we expect $qs<1$. 
We found that $qs$=2/3 gives a good fit to most data points (Figure \ref{fig:migincmass}).  We want to emphasize that $qs$ is the fudge factor which we use to represent complicated physical processes. In reality, $qs$ could vary with both disc and planet parameters, and it can be different between migration rates and inclination damping rates. The two noticeable outliers are the high migration rates for planets with very deep gaps. Thus, something else besides the dynamical friction is pushing the planet inwards. We suspect that the high migration rates are related to the precession of the inner discs in these two cases. An analytical theory which takes into account both the non-uniform gap density
profile and the precession of the inner disc is needed in future.

\subsection{Gap Opening and Disc Breaking}

\begin{figure} 
\centering
\includegraphics[trim=0cm 0cm 0cm 0cm, width=0.5\textwidth]{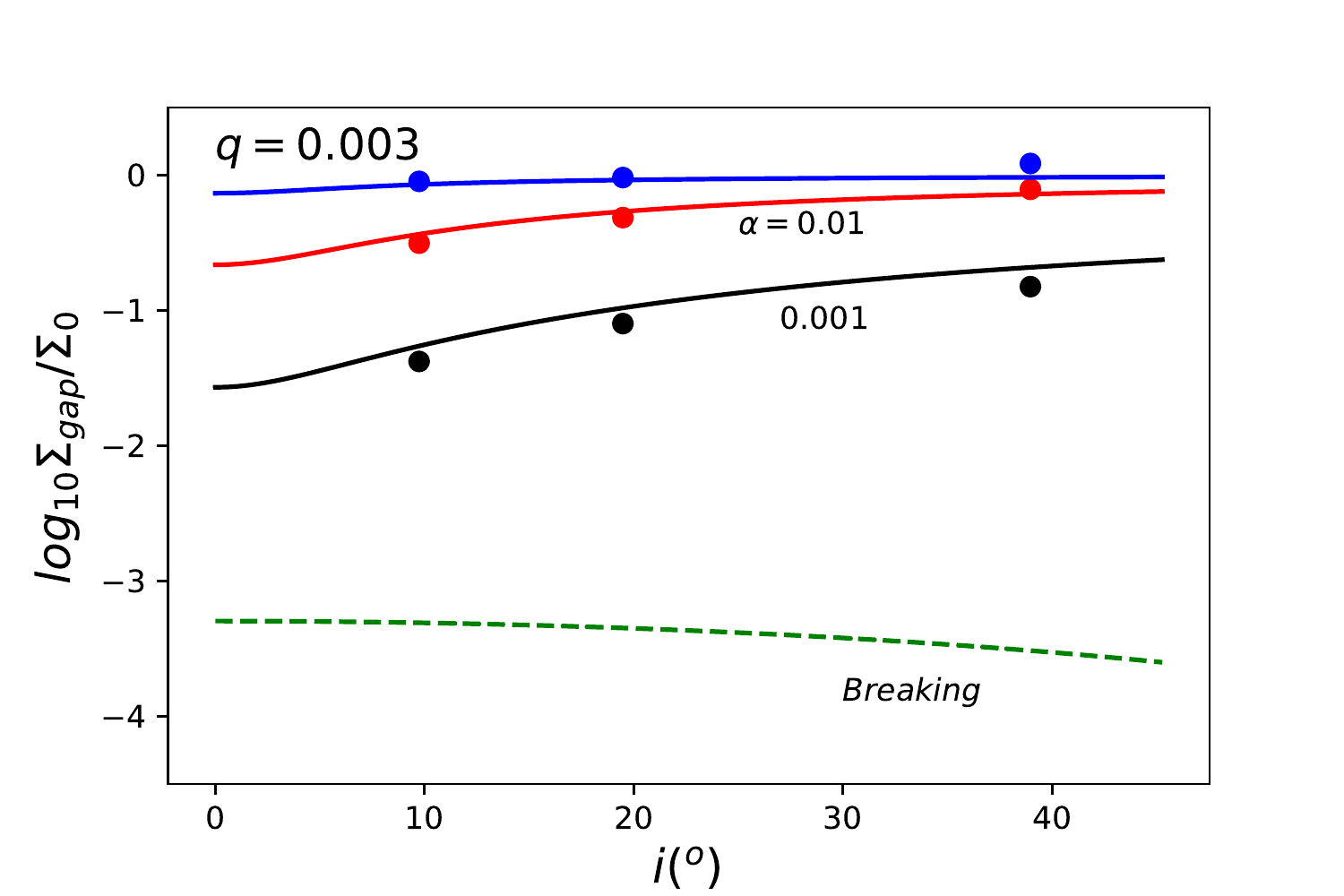} \\
\includegraphics[trim=0cm 0cm 0cm 0cm, width=0.5\textwidth]{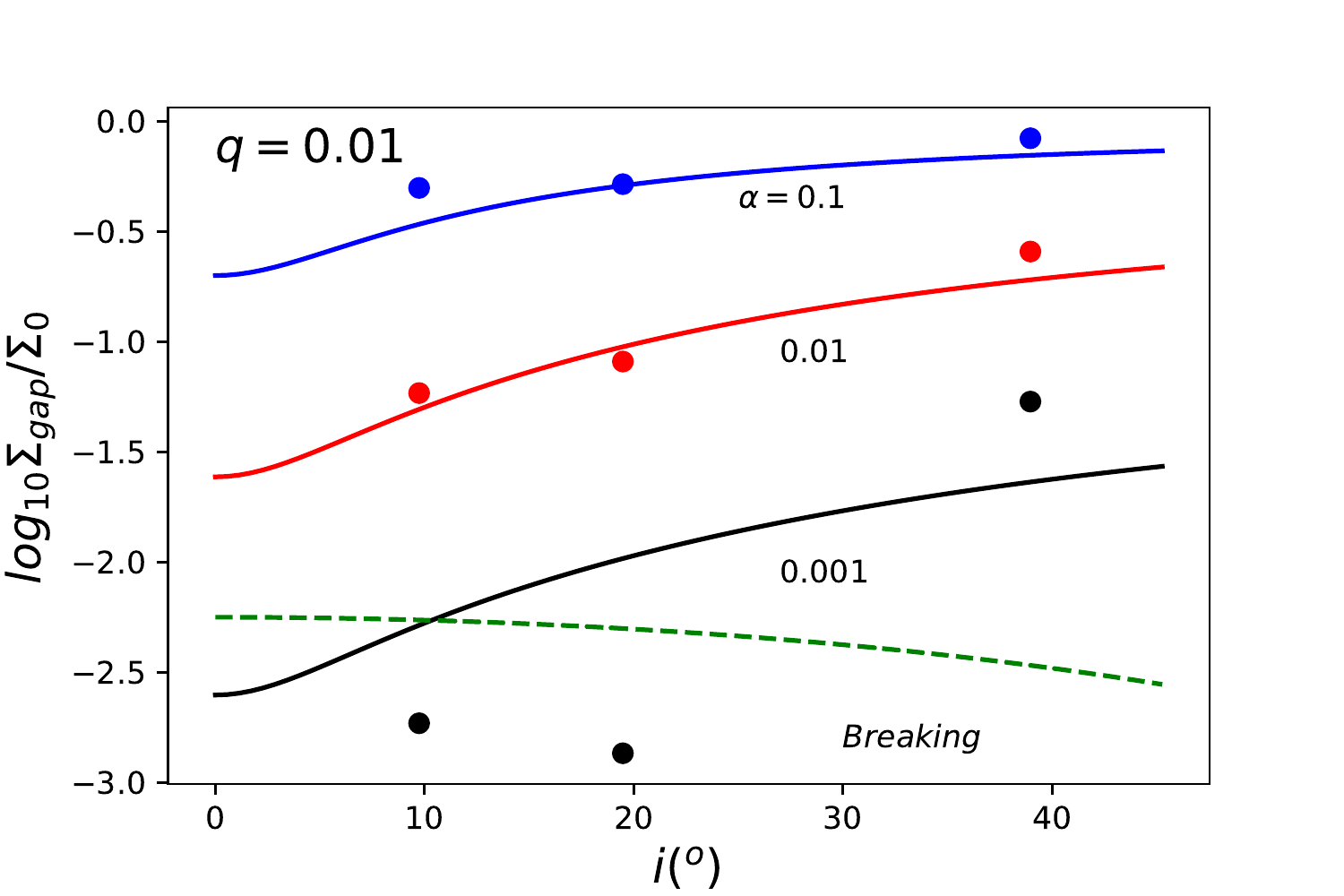} 
\caption{The gap depth with respect to the planet inclination for a $q=0.003$ (upper panel) and $q=0.01$ (lower panel) 
planet in the disc. The dots are depths measured from the numerical simulations, while the solid curves
are from the analytical formula (Equation \ref{eq:depthwidth}) and \ref{eq:newK}. The green dashed curves 
are the disc breaking condition from Equation \ref{eq:breaking}.}
\vspace{-0.1 cm} \label{fig:gapbreaking}
\end{figure}

After studying the planet's orbital evolution, we study the disc that is under the influence of the planet.
We use simulation results to test our analytical estimate on the gap depth and the disc breaking condition. 
Analytical theory (Equation \ref{eq:depthwidth}) suggests that the gap depth and width for a misaligned planet are similar
to those for a coplanar planet but with a modified planet mass (Equation \ref{eq:newq} and \ref{eq:newK}). In Figure \ref{fig:gapbreaking},
we plot both the measured gap depth from simulations and the prediction from the analytical formula (Equation \ref{eq:newK}). Good agreements
are found for both $q=0.003$ and $q=0.01$ cases, as long as the disc is not undergoing disc breaking (\texttt{P10I10AM3}, \texttt{P10I19AM3} have disc breaking.).

To test our disc breaking condition (Equation \ref{eq:breaking}), we plot Equation \ref{eq:breaking} as dashed curves in Figure \ref{fig:gapbreaking}.
Whenever the dashed curve is larger than the
gap depth curve, the disc should undergo breaking. With a $q$=0.003 planet, the gap is not deep enough
for disc breaking even with $\alpha=10^{-3}$. On the other hand, if the planet has $q$=0.01 and a low inclination, the 
induced gap in the $\alpha=10^{-3}$ disc is deep enough for the disc breaking. 
The fact that the discs in \texttt{P10I10AM3} and \texttt{P10I19AM3} undergo breaking is consistent  with our breaking condition.

\begin{figure} 
\centering
\includegraphics[trim=0cm 0cm 0cm 0cm, width=0.5\textwidth]{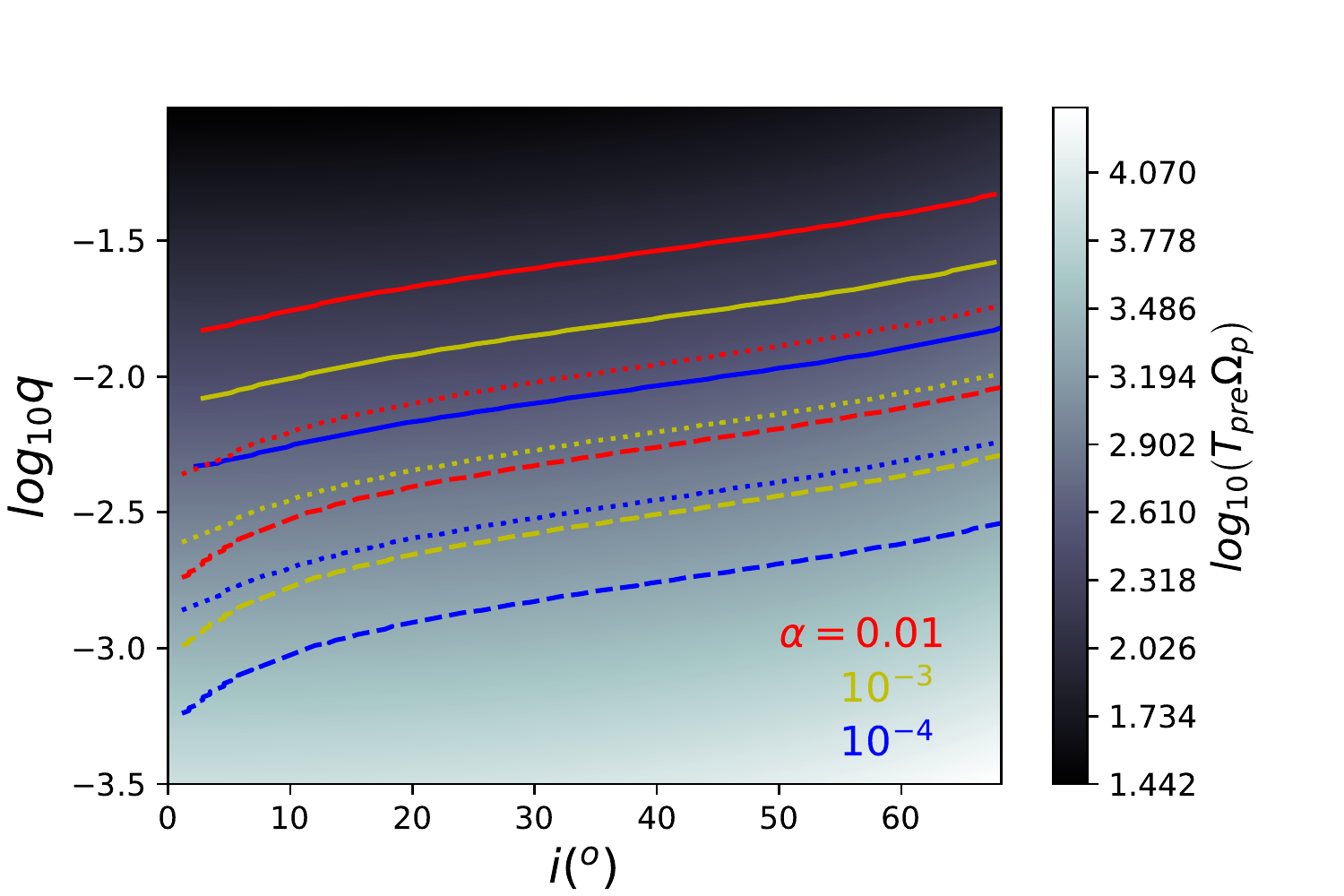} 
\caption{ The disc-breaking planet mass for different values of $\alpha$ (curves with different colors) and different values of $H/R$ (0.1: solid
curves, 0.05: dotted curves, 0.03: dashed curves). The gray map shows the disc precession timescale. }
\vspace{-0.1 cm} \label{fig:precessionmass}
\end{figure}

With the gap depth formula (Equation \ref{eq:depthwidth}) and the disc breaking condition (Equation \ref{eq:breaking}),
we can calculate how massive the planet needs to be in order to break the disc and cause differential precession and large misalignment. 
For given disc parameters ($H/R$ and $\alpha$) and a given planet inclination ($i_p$), we gradually increase the planet mass until
the gap depth (Equation \ref{eq:depthwidth}) reaches the disc breaking condition (Equation \ref{eq:breaking}). We define this planet
mass as the disc-breaking planet mass. Planets with masses larger than this mass can cause disc breaking, while planets with less mass
cannot break the disc and there will be a constant relative twist between the inner and outer discs.
Figure \ref{fig:precessionmass} shows the disc-breaking planet mass for different values of $\alpha$ (different colors) and different values of $H/R$ (0.1: solid
curves, 0.05: dotted curves, 0.03: dashed curves). We can see that only in extreme conditions, e.g. $H/R\lesssim$0.05 and $\alpha\lesssim10^{-4}$, a Jupiter mass planet can break the disc. This figure also predicts that, in our thin disc simulation 
\texttt{THINP10I39AM3}, the thin disc with $H/R$=0.05 and $\alpha=10^{-3}$  (the yellow dotted curve) should break by the   $39^o$ inclined $10\,M_{J}$ planet. We confirm this by checking the \texttt{THINP10I39AM3} simulation directly.

\section{Observational Signatures}
The observational signatures of warped discs \citep{Facchini2014, Juhasz2017} and broken discs \citep{Facchini2018} have been studied recently. Although these calculations use circumbinary disc simulations, most of their results can also be applied to warped and broken discs induced by the planet, as shown in this section. We will also highlight some differences between the observational signatures of the broken circumbinary discs and the broken discs induced by the planet.

We calculate the near-IR scattered light images and (sub-)mm velocity channel maps for two simulations (\texttt{P10I19A0} and \texttt{THINP10I39AM3}) where the gap is deep enough for the disc to break. The inner and outer discs thus have significant misalignment due to their different precession frequencies. 
We use RADMC-3D\footnote{RADMC-3D is an open code of radiative transfer calculations. The code is
available online:
http://www.ita.uni-heidelberg.de/~dullemond/software/radmc-3d/.} for the 3-D radiative transfer calculations, following the similar procedures as in \cite{Arzamasskiy2018}. A spherical mesh is adopted.
Since one level of mesh refinement is used in our hydrodynamical simulations,  we coarsen the fine mesh to generate a mesh with the uniform resolution. To scale the quantities in our simulations, the length unit is set to be 100 au, and the initial gas surface density at 100 au is set to be either 1 g cm$^{-2}$ or 100 g cm$^{-2}$. The central star is assumed to be a Herbig Ae/Be star with $T_{eff}=6810$ K, $R_*$=1.4 $R_{\odot}$, and $M_*$=1.7 $M_{\odot}$. The initial dust to gas mass ratio is 1/100, and we have assumed that 10\% of dust mass is in small grains which determine the disc temperature structure and scatter light at near-IR. The dust opacity is calculated using Mie theory for Magnesium-iron silicates \citep{Dorschner1995} assuming a size of 0.1 $\mu$m.  

The radiative transfer calculation with RADMC-3D includes 3 steps. First, the temperature structure of the disc is calculated. Second, using this temperature structure, the near-IR scattered light images are generated including full treatment of light polarization. The Stokes parameters (I, Q, U, V) are computed and the polarized intensity is $\sqrt{Q^2+U^2}$ (V=0 due to linear polarization from dust scattering). 
Third, assuming that the number density ratio between $^{13}$CO and H$_2$ is 1.75$\times10^{-6}$, the channel maps for the $^{13}$CO (3-2) line are generated. 

The images presented below assume a geometry that we are viewing towards the disc along the polar direction in the simulation (so the simulation's $z$-axis is pointing towards us). Thus,
the orbit of the planet is on the plane of sky. The $x$-axis in our simulations points to the right direction in the images and the $y$-axis in the simulations points to the up direction in the images. 

\subsection{The Near-IR Scattered Light Image}
\begin{figure*} 
\centering
\includegraphics[trim=0cm 1.3cm 0cm 1cm, width=0.8\textwidth]{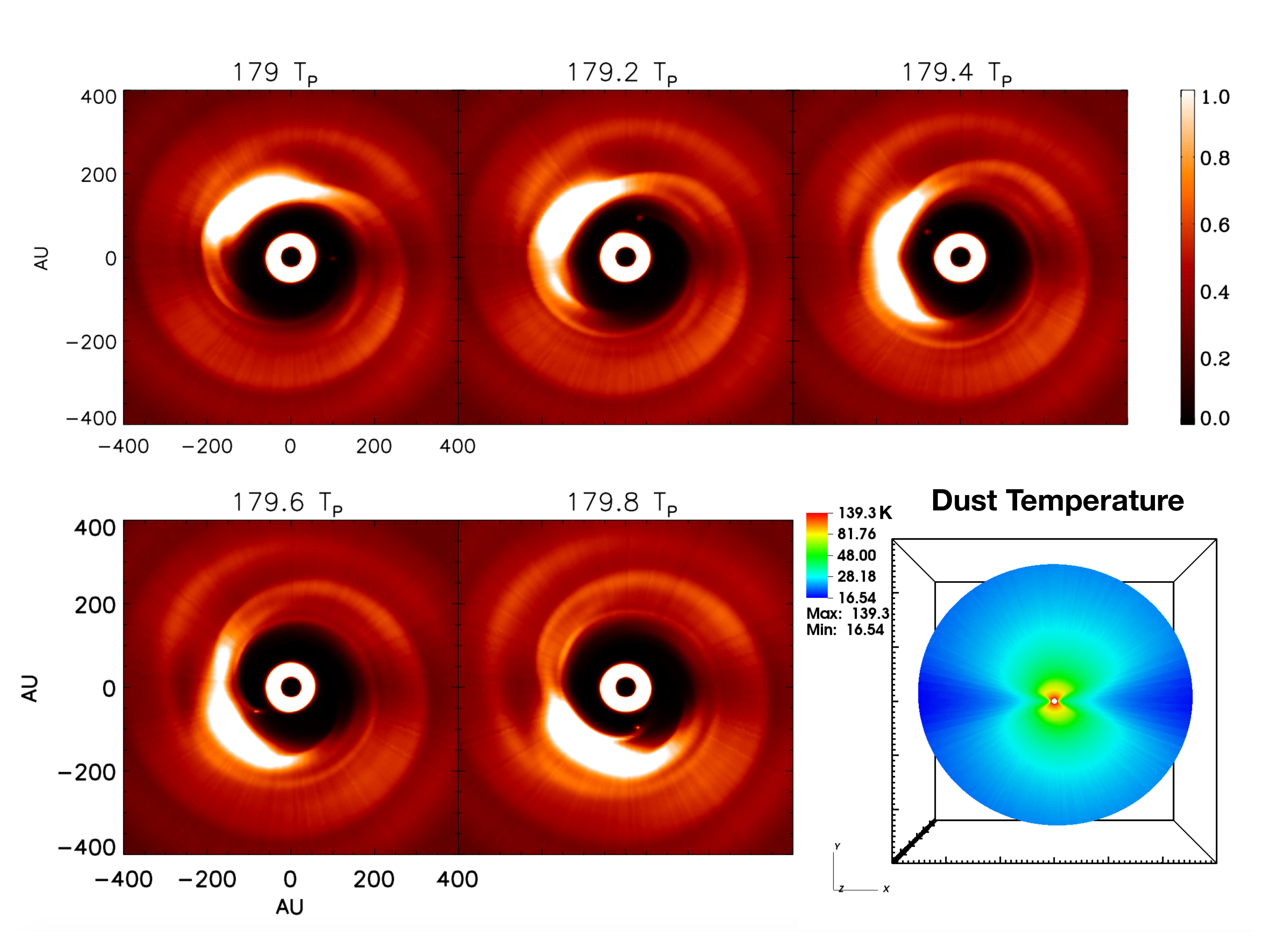} 
\caption{The near-IR scattered light images for \texttt{P10I19A0} with $\Sigma_0=1$ g cm$^{-2}$ at different times. We can see the gap edge vortex is moving across the shadow from the inner disc. The lower right panel shows the disc temperature at the midplane. }
\vspace{-0.1 cm} \label{fig:scatvortex}
\end{figure*}

The near-IR scattered light images for \texttt{P10I19A0} with $\Sigma_0=1$ g cm$^{-2}$ are presented in Figure \ref{fig:scatvortex}.  The intensity has been multiplied by $R^2$ in the image. The colorbar is linear with the maximum chosen to highlight the disc features (the brightest color represents a normalized intensity that is a factor of 5 smaller than the maximum normalized intensity in the image). Since this is an inviscid simulation, we clearly see the gap edge vortex in the images. On the other hand, we also see a dark lane in the horizontal direction in the image. This dark lane is along the nodal lines between the inner and outer discs, and thus is due to the shadow from the inner inclined disc. Since the vortex orbits around the central star at the local Keplerian speed, we can see that the vortex is traveling across the dark lane in these snapshots. When the vortex and the dark lane overlap (T=179.4 $T_p$), the vortex looks like it splits into two vortices.  After the vortex travels past the dark lane it looks like one vortex again. This is consistent with HD 142527 where the submm vortex seems to split into two parts \citep{Casassus2015} at the dark lane of the near-IR scattered light image \citep{Marino2015}. Although the split of the vortex in HD 142527 is shown at (sub-)mm wavelengths where the dust thermal emission dominates, it can still be related to the shadow due to the decrease of the dust temperature at the shadow lane. The lower right panel of Figure \ref{fig:scatvortex} shows the dust temperature at the disc midplane. Clearly, the casting shadow lowers the disc temperature there, which is also seen in broken circumbinary disc simulations \citep{Juhasz2017,Facchini2018}.

\begin{figure*} 
\centering
\includegraphics[trim=0cm 0.5cm 0cm 1.8cm, clip, width=0.8\textwidth]{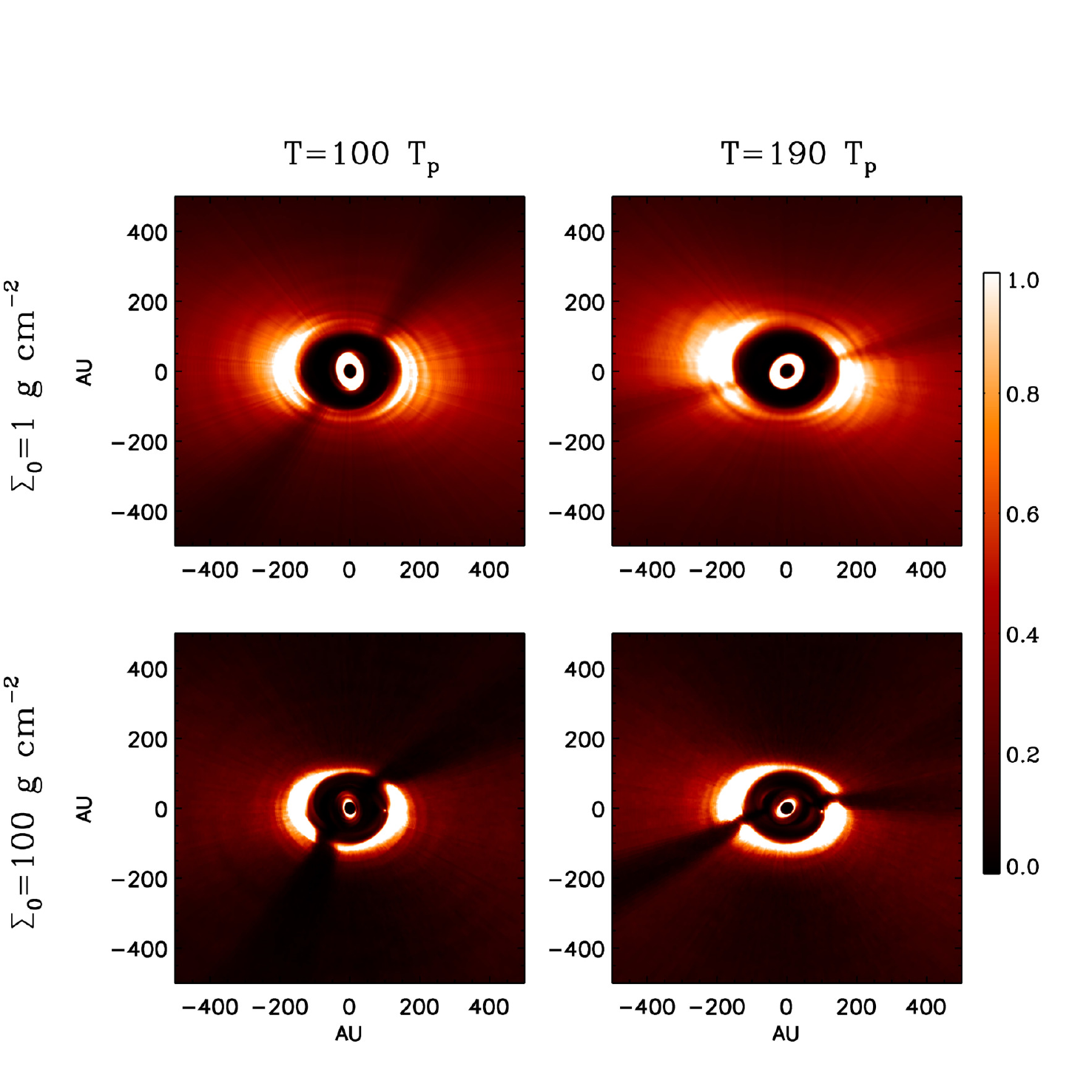} 
\caption{The near-IR scattered light images for \texttt{THINP10I39AM3} with $\Sigma_0=1$ g cm$^{-2}$ (upper panels) and 100 g cm$^{-2}$ (lower panels) at T=100 $T_p$ (left panels) and 190 $T_p$ (right panels). While the inner disc precesses over $60^o$ between the left and right panels, the shadow only rotates for 30 $^o$ as discussed in the text. }
\vspace{-0.1 cm} \label{fig:scatpaper}
\end{figure*}

\begin{figure} 
\centering
\includegraphics[trim=10cm 10cm 8cm 5cm, clip, width=0.5\textwidth]{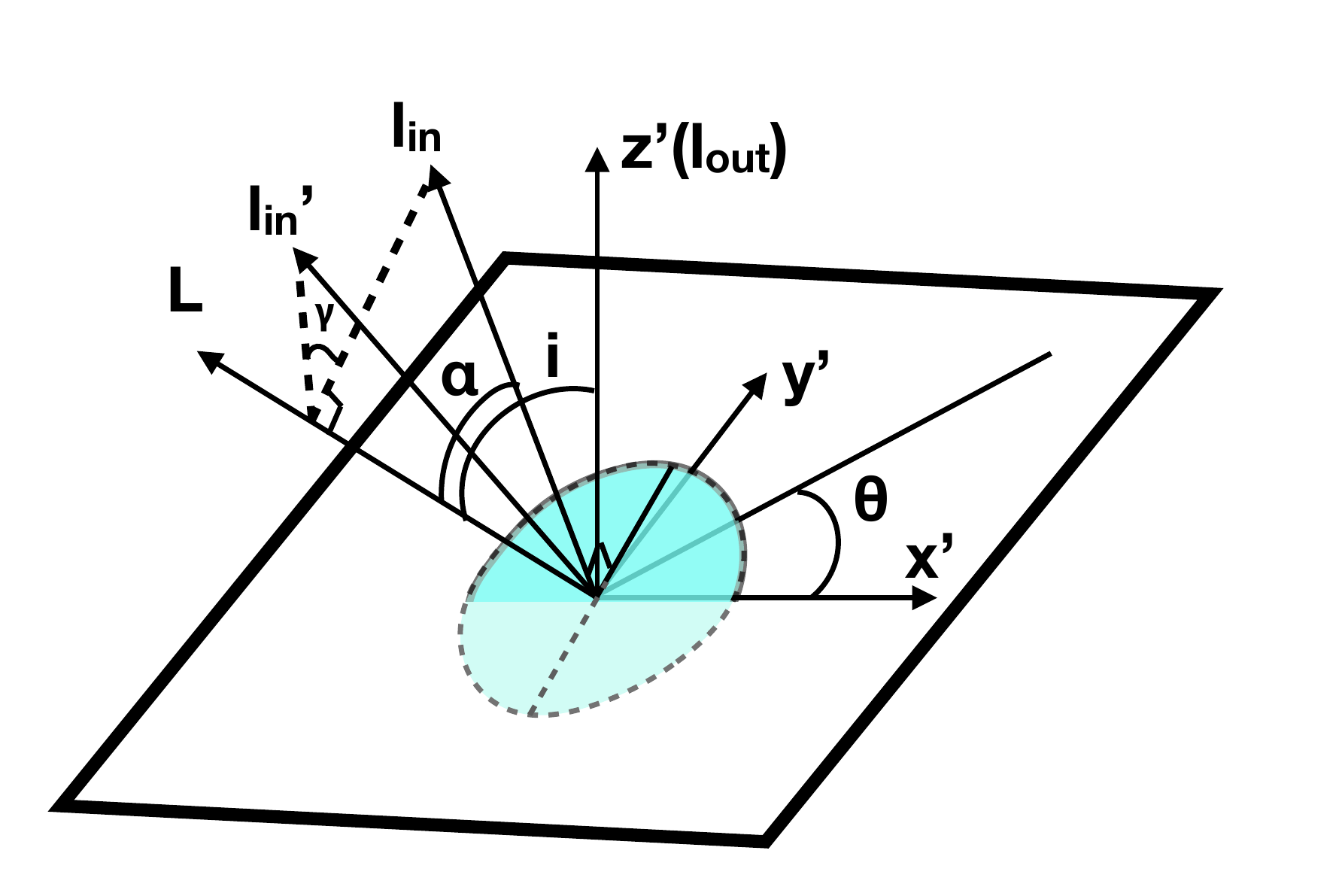} 
\caption{ The geometry of the inclined disc and the casting shadow. The inner disc with the angular momentum vector of ${\bf l_{in}}$  precesses around the total angular momentum vector ${\bf L}$ for the angle of $\gamma$. The line of nodes, which is also the direction of the casting shadow has an angle of $\theta$ from the $x'$-axis. Both ${\bf L}$ and the intial ${\bf l_{in}}$ are pointing to the negative $y'$ direction in the $y'-z'$ plane. }
\vspace{-0.1 cm} \label{fig:geometry}
\end{figure}

\begin{figure*} 
\centering
\includegraphics[trim=0cm 1.1cm 0cm 1.3cm, clip, width=1.\textwidth]{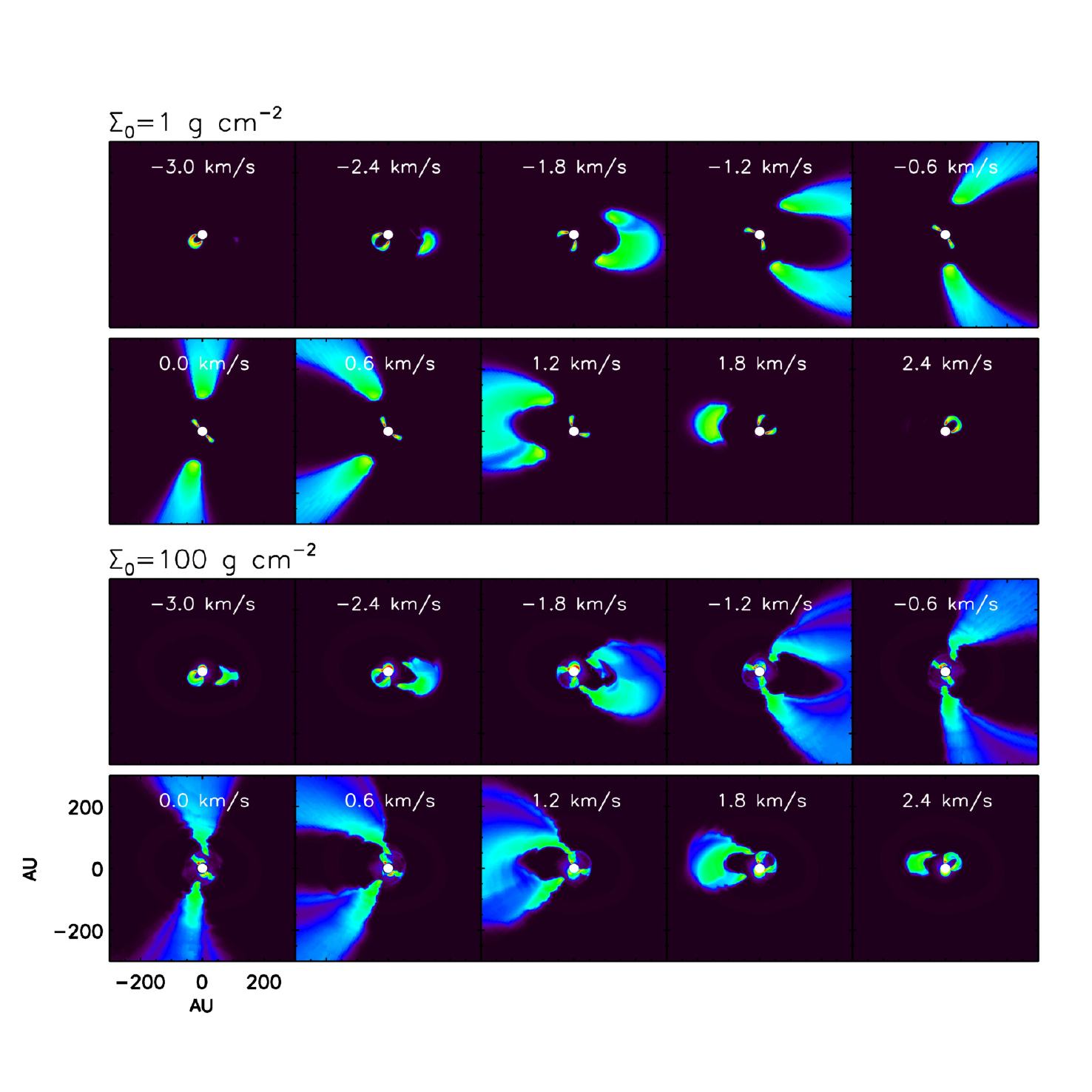}
\caption{The $^{13}$CO (3-2) channel maps for \texttt{THINP10I39AM3} at T=190 $T_p$. The upper panel block is with $\Sigma_0=1$ g cm$^{-2}$ and the lower panel block is with $\Sigma_0=100$ g cm$^{-2}$. }
\vspace{-0.1 cm} \label{fig:channel}
\end{figure*}

\begin{figure*} 
\centering
\includegraphics[trim=0cm 0.8cm 0cm 1.6cm, clip, width=1.\textwidth]{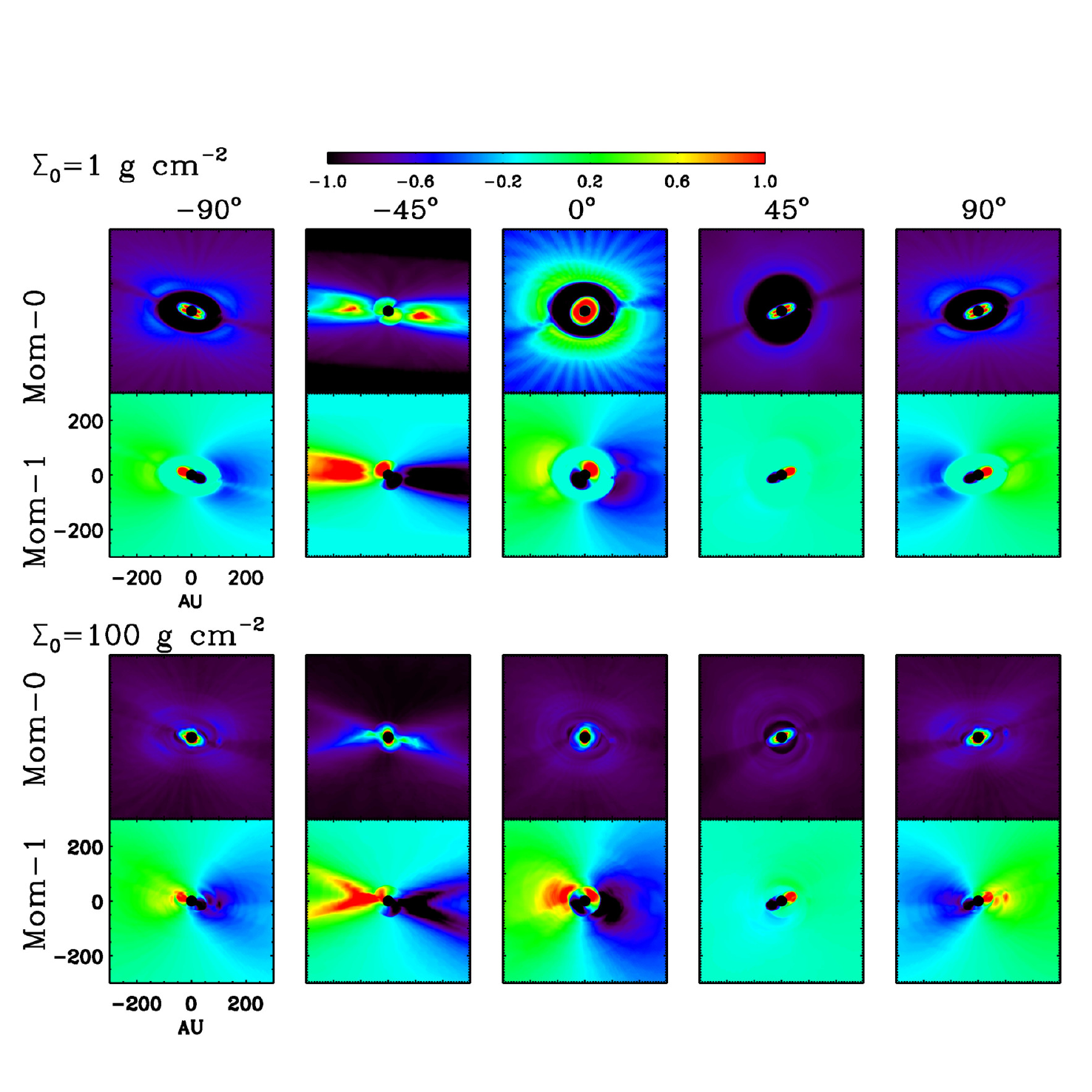}
\caption{The moment 0 and moment 1 maps for  $^{13}$CO  (3-2) for \texttt{THINP10I39AM3} with $\Sigma_{0}=1\, g\, cm^{-2}$ (the upper panel block) or $\Sigma_{0}=100\, g\, cm^{-2}$ (the lower panel block)  when the disc is viewed at different viewing angles. The inclination of the planet's angular momentum vector with respect to our line of sight is labeled above the images.   }
\vspace{-0.2 cm} \label{fig:momentum}
\end{figure*}

The near-IR scattered light images for \texttt{THINP10I39AM3} (Figure \ref{fig:scatpaper}) are quite different from those for \texttt{P10I19A0}. First, due to the finite viscosity with $\alpha=10^{-3}$, vortices do not show up at the gap edge. This is consistent with simulations having coplanar planets in discs \citep{Fu2014,ZhuStone2014,Hammer2017}. Second, the shadows are much sharper. This is because, in \texttt{THINP10I39AM3}, the inner disc is both thinner and more misaligned. The larger misalignment is because the planet in \texttt{THINP10I39AM3} has a higher inclination   (39$^o$ compared with 19$^o$ in \texttt{P10I19A0}) so that the differential precession can lead to a much larger misalignment between the inner and outer discs ($\sim 60^0$ in Figure \ref{fig:scatpaper} compared with $\sim 35^o$ in Figure \ref{fig:scatvortex}).

We also increase the disc surface density by a factor of 100 (lower panels of Figure \ref{fig:scatpaper}) to explore how the disc surface density affects the disc shadow.
Two effects can be seen in the figure. First, 
when the disc surface density is higher, the scattering surface at the inner disc is higher so that the inner disc casts a wider shadow on the outer disc. Second, when the disc surface density is higher, the scattering surface at the outer disc is also higher so that the shadow is cast onto a bowl shaped flaring surface instead of a flat surface close to the midplane.
As shown in the bottom panels of Figure \ref{fig:scatpaper}, the two dark lanes concave towards one direction tracing the scattering surface. Thus, we can use the dark lane positions to reconstruct the scattering surface in the disc.

Another interesting phenomenon we observe is that the shadow moves at a slower speed than the inner disc precession speed. By checking the simulation directly, we find that
the inner disc's tilt angle is 60$^o$ at $T=100\, T_{p}$ and 120$^o$ at $T=190\, T_{p}$. Although the inner disc's tilt angle changes by 60$^o$, the shadow only rotates $\sim 30^o$ by comparing the left and right panels in Figure \ref{fig:scatpaper}. To understand this difference, Figure \ref{fig:geometry} shows the geometry of the inner and outer discs. We assign the outer disc midplane as the $x'-y'$ plane to be different from the planet's orbital plane that is denoted as the $x-y$ plane. The total angular momentum vector ${\bf L}$ is at the $y'-z'$ plane and is pointing to the negative $y'$ direction. The tilt angle between ${\bf L}$ and the $z'$ axis is $i$. The inner disc's angular momentum vector is ${\bf l_{in}}$ and is also at the $y'-z'$ plane initially. The angle between ${\bf L}$ and ${\bf l_{in}}$ is $\alpha$. After some time, ${\bf l_{in}}$ precesses around ${\bf L}$ for an angle of $\gamma$ to ${\bf l_{in}'}$. We assume that the outer disc does not precess during this short period of time. The line of nodes, which is also the direction of the casting shadow, has an angle of $\theta$ from the $x'$-axis in the $x'-y'$ plane. The relationship between all these angles is
\begin{equation}
   {\rm tg} \theta = \frac{{\rm sin} \alpha \cdot {\rm sin} \gamma}{{\rm sin}\alpha \cdot {\rm cos}\gamma \cdot {\rm cos} i-{\rm cos}\alpha\cdot {\rm sin}i}\,.\label{eq:shadowrate1}
\end{equation}
In our simulation setup, the inner and outer discs are coplanar initially. Thus ${\bf l_{in}}$  and the $z'$-axis overlap and $\alpha=i$. So the equation is simplified to
\begin{equation}
    {\rm tg} \theta = \frac{ {\rm sin} \gamma}{{\rm cos} i \cdot({\rm cos}\gamma -1)}\,.
\end{equation}
If we plug in $i=-39^o$ and $\gamma=-60^o$ at $T=100\, T_{p}$, $\theta$ is $66^o$. If we plug in $i=-39^o$ and $\gamma=-120^o$ at $T=190\, T_{p}$, $\theta$ is $37^o$. Considering that $\theta$ is the angle from the positive $x'$-axis in the counterclockwise direction, these angles are consistent with the dark lanes in Figure \ref{fig:scatpaper}. Thus, while the inner disc  precesses for $60^o$, the shadow only precesses for $\sim 30^o$. We can use Equation \ref{eq:shadowrate1} to explore several extreme scenarios. If the total angular momentum vector ${\bf L}$ is along the outer disc angular momentum vector ($z'$-axis), we have $i=0$ and thus ${\rm tg}\theta={\rm tg} \gamma$. The shadow rotates at the same frequency as the inner disc precession. For the other extreme, if $\alpha=i=90^o$, $\theta$ is always 90$^o$ and the shadow never moves.   These indicate that the rotation of shadow not only depends on the inner disc precession but also depends on the geometry of the system. The shadow may or may not move at the same frequency as the disc precession frequency.

Our derivation is similar to \cite{Min2017} where they have pointed out the position angle of the shadow is different from the position angle of the inner disc, except that we choose a different coordinate system that is aligned with the outer disc. This coordinate system is motivated by the disc's nodal precession. It not only simplifies the calculation, but also makes it much easier to study the time variability of the shadow due to the disc precession.
 
We can relate the movement of the shadow with the inner disc's precession by doing time derivative for Equation \ref{eq:shadowrate1}. We then have
\begin{equation}
    \frac{d\theta}{dt}=\frac{{\rm sin}^2\alpha \cdot {\rm cos} i-{\rm sin}\alpha\cdot{\rm cos}\alpha\cdot{\rm sin}i\cdot{\rm cos}\gamma}{\left({\rm sin}\alpha \cdot {\rm cos}\gamma \cdot {\rm cos} i-{\rm cos}\alpha\cdot {\rm sin}i\right)^2+{\rm sin}^2\alpha \cdot{\rm sin}^2\gamma}\frac{d\gamma}{dt}\,,\label{eq:dthetadt}
\end{equation}
where  $d\theta/dt$ is the rotational rate of the shadow while $d\gamma/dt$ is the precession rate of the inner disc. In our simulation setup with $\alpha=i$, we have
\begin{equation}
    \frac{d\theta}{dt}=\frac{{\rm cos}i\cdot(1-{\rm cos}\gamma)}{ {\rm cos}^2 i\cdot({\rm cos}\gamma-1)^2+{\rm sin}^2 \gamma}\frac{d\gamma}{dt}\,.
\end{equation}
Thus, the shadow can move faster or slower than the inner disc precession, depending on both $i$ and $\gamma$.

\subsection{CO Channel Maps}

The $^{13}$CO channel maps for \texttt{THINP10I39AM3} are presented in Figure \ref{fig:channel}. Initially, both the inner  and outer discs are oriented in a way that the right side in the image is  blue-shifted while the left side is  red-shifted.
At $T=190$ $T_p$, the inner disc precesses around the $z$-axis (the $z$-axis is pointing to us along our line of sight) for 120$^o$ in the clockwise direction (or retrogradely compared with the planet's orbital motion). Thus, the channel maps of the inner disc starts at $\sim$120$^o$ away from the positive $x$-axis in the clockwise direction, while the outer disc still starts at the positive $x$-axis. Besides the misalignment between the inner and outer discs, the shadow at $\theta\sim$30$^o$ and $\sim$-150$^o$ is also visible from the channel maps. This is due to the decrease of the disc temperature within the shadow. One noticeable difference between the $\Sigma_0=1$ g cm$^{-2}$ and 100 g cm$^{-2}$ cases is that, for the latter case, the $^{13}$CO emission surface is high in the atmosphere where it is hotter than the midplane so that we can see both the front and back sides of the disc \citep{Rosenfeld2013}. 

The moment 0 and 1 maps that are derived from the channel maps in Figure \ref{fig:channel} are shown in the middle column of Figure \ref{fig:momentum}. We see less emission at the shadow lane in the moment 0 maps due to the lower temperature there. The different rotational directions between the inner and outer discs are apparent in the moment 1 maps. Even if we view this system at different angles (the other columns in Figure \ref{fig:momentum}), we can still clearly see the different disc orientations between the inner and outer discs. In the -45$^o$ maps, we are viewing the outer disc almost edge on, while viewing the inner disc almost face on. In the 45$^o$ maps, we are viewing the inner disc edge on while viewing the outer disc face on. If the molecular tracer is optically thin, we can clearly see a wide gap between the inner and outer discs, which is different from disc breaking in circumbinary discs. On the other hand, when the disc surface density is high (the bottom panel block in Figure \ref{fig:momentum}) or the line is more optically thick (e.g. $^{12}$CO lines), we may not see the gap between the inner and outer discs. But we can still see that the velocity structure is twisted in the moment 1 maps. Overall, unlike the near-IR scattered light images, the CO (sub-)mm moment maps provide direct information on the  orientation of both the inner and outer discs, which can be quite helpful for probing the potential companion.

\section{Discussion}

\subsection{Various Timescales}

\begin{figure} 
\centering
\includegraphics[trim=0cm 0.5cm 0cm 0cm, width=0.5\textwidth]{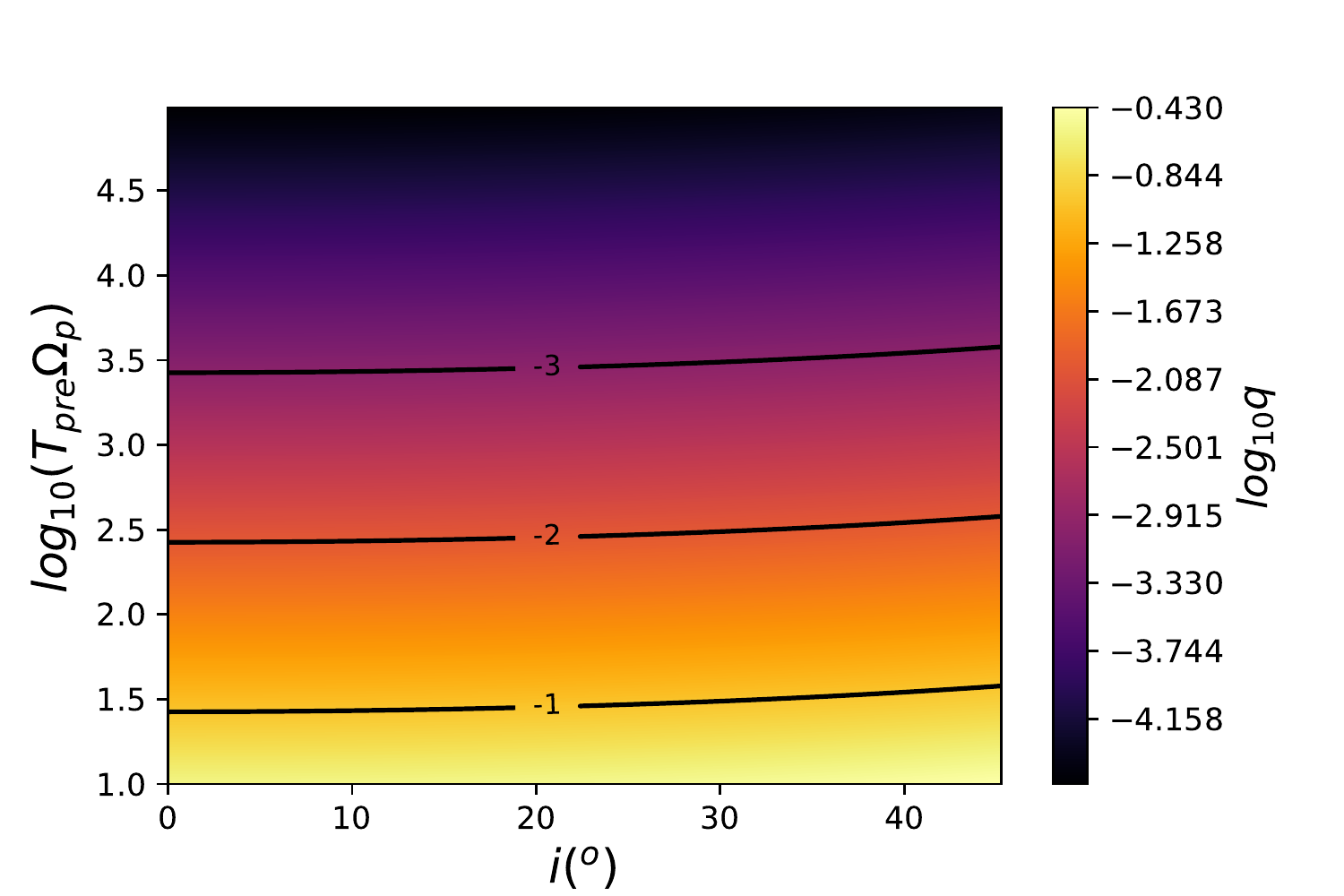} 
\caption{The relationship between the disc precession timescale, the planet inclination, and the planet mass. The $y$-axis is the precession timescale and the color scale is the planet mass. The three solid curves represent $q=10^{-3}$, $q=10^{-2}$, and $q=10^{-1}$.}
\vspace{-0.2 cm} \label{fig:precessiontime}
\end{figure}

If the planet is massive enough to break the disc, the inner disc will precess at the precession timescale (Equation \ref{eq:precessionrate}).
Both Figures \ref{fig:precessionmass} and \ref{fig:precessiontime} show
the relationship between the disc precession timescale, the planet inclination, and the planet mass. In  Figure \ref{fig:precessionmass},
the $y$-axis is the planet mass while the color scale is the precession timescale. In Figure \ref{fig:precessiontime}, the $y$-axis
is the precession timescale and the color scale is the planet mass. Combing these two figures, we can easily constrain the disc precession timescale if the disc is known to be broken. For example, for a broken disc with $\alpha=10^{-3}$ and $H/R=0.03$, Figure \ref{fig:precessionmass} tells us
that $q$ has to be larger than $10^{-3}$ to break the disc. Then, Figure \ref{fig:precessiontime} or Equation \ref{eq:precessionrate} informs us that with $q>10^{-3}$ the disc precession timescale is shorter than 3000/$\Omega_p$.

\begin{figure} 
\centering
\includegraphics[trim=0cm 1.cm 0cm 0cm, width=0.5\textwidth]{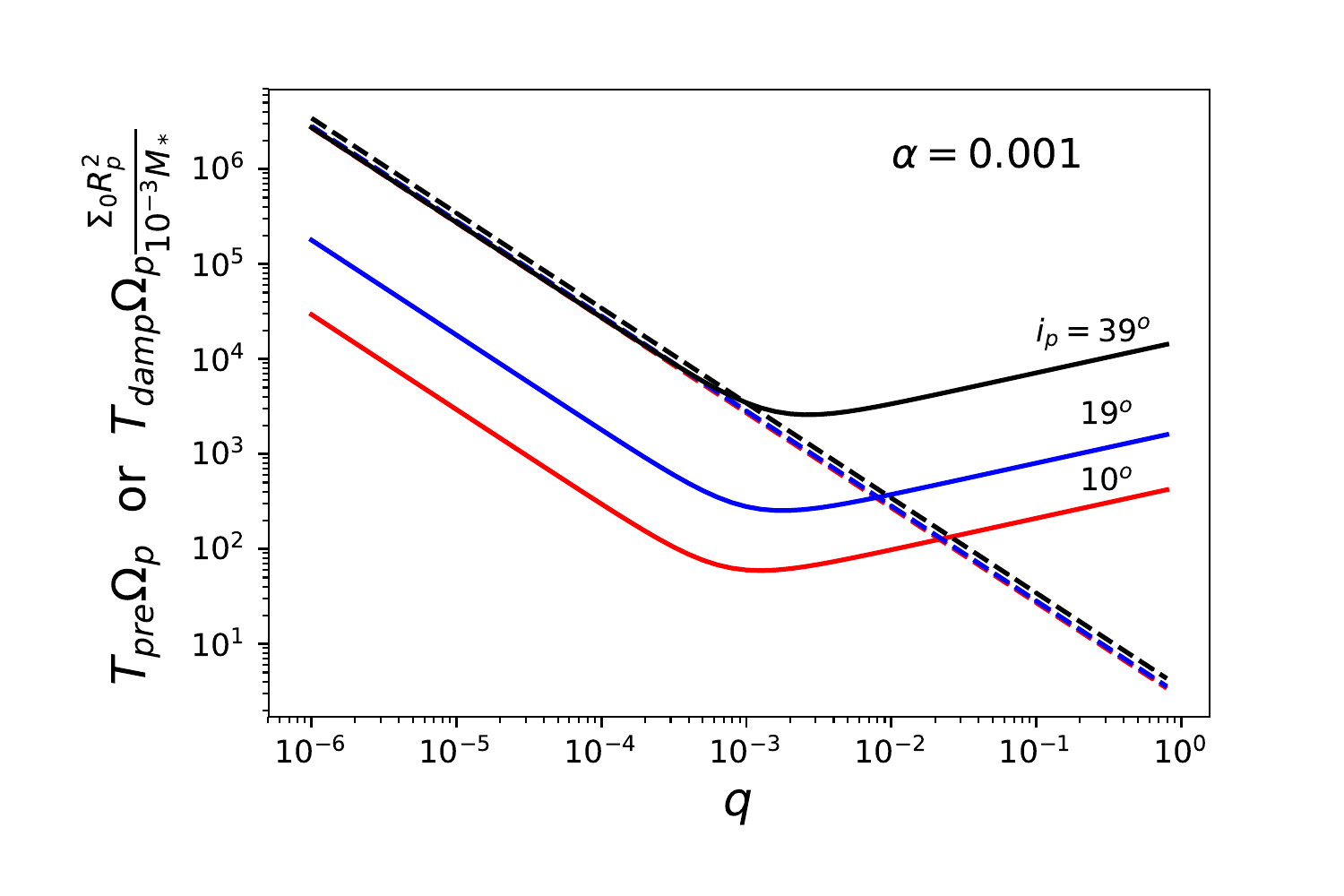} 
\caption{Solid curves: the inclination damping timescales for planets with different inclinations. The $\alpha$ in the disc is $10^{-3}$. After gap opening,
the inclination damping timescale becomes longer. The inclination damping timescale has been scaled with
$\Sigma_0R_p^2/(0.001M_*)$. For comparison, the  dashed lines are the disc precession timescales  for these three inclinations (the corresponding $y$-axis is $T_{pre}\Omega_p$). }
\vspace{-0.1 cm} \label{fig:damptime}
\end{figure}

Although the disc-breaking planet can cause the inner disc to precess (\S 4.3), the planet's inclination will be damped at the same time (\S 4.2).
If the inclination damping timescale is shorter than the disc precession timescale, 
we should not expect misaligned inner and outer discs since the disc will not have
enough time to precess before the planet becomes coplanar. Figure \ref{fig:damptime} shows both the inclination damping
timescale (Equation \ref{eq:tincgap}) and the disc precession timescale (Equation \ref{eq:precessionrate}). The precession timescale
has a very weak dependence on the planet inclination (cos $i_p$), while the inclination damping timescale sensitively depends on
the planet inclination. Figure \ref{fig:damptime} shows that the inclination damping timescale decreases until a gap starts to form, and then
the inclination damping timescale increases. If $\Sigma_0 R_{p}^2/M_*$=0.001 and $\alpha=10^{-3}$,
the inclination damping timescale for a $i=39^o$ planet is comparable or longer than the precession timescale, and significant misalignment between the inner
and outer discs are possible. For less inclined planets, only massive planets ($q\gtrsim 0.01$) can cause significant misalignment before the planet becomes
coplanar. When the disc is more massive or $\alpha$ is larger, the inclination damping timescale gets shorter and
an even more massive planet is needed to cause misalignment between the inner and outer discs. On the other hand, if we can constrain that the planet in the misaligned disc is not very massive (e.g. 1 M$_J$), we can infer that the disc $\alpha$ is not large (e.g. $\alpha\lesssim 10^{-3}$) or the disc mass is small (e.g. $\Sigma_0 R_{p}^2/M_*\lesssim 10^{-3}$). 

\begin{figure} 
\centering
\includegraphics[trim=0cm 1.cm 0cm 0cm, width=0.5\textwidth]{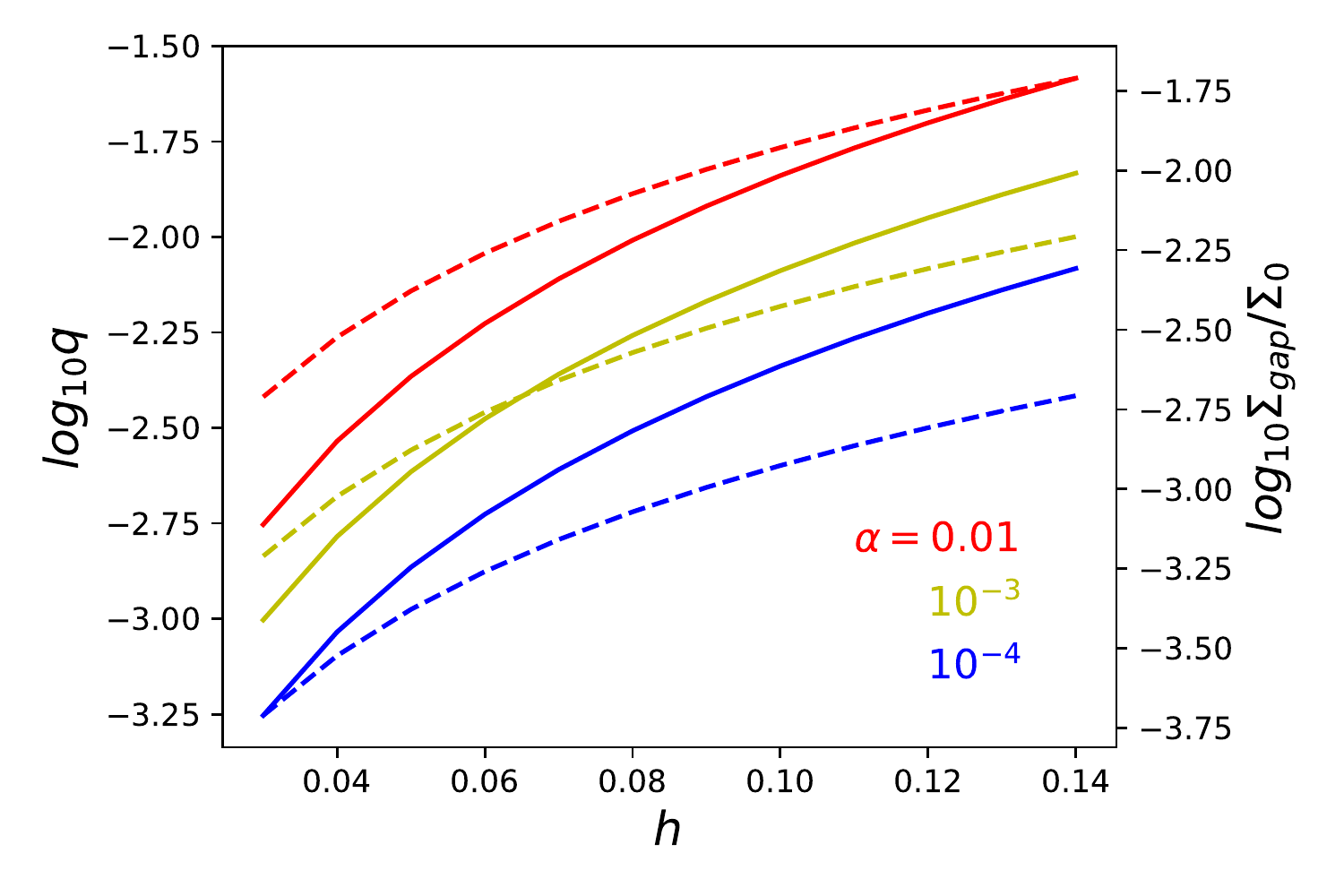} 
\caption{The minimum planet mass which can break the disc (solid curves with the left axis) and the corresponding gap depth (dashed curves with the right axis) for discs having different $h$. Different colors represent discs with different viscosity coefficients ($\alpha$). }
\vspace{-0.1 cm} \label{fig:breakingmass}
\end{figure}

\subsection{Inferring Planet Properties From Observations}
Based on observations, the geometry of the broken discs can be nicely constrained by studying its near-IR images and (sub-)mm molecular line channel maps (\S 5). We want to use this broken disc geometry to constrain the mass and inclination of the potential planet in the disk.

The planet's inclination is relatively easy to derive if the inner and outer discs only experience the nodal precession around the system's angular momentum vector. Under this condition (which can be violated in some cases as discussed in \S 6.4), the angle between the angular momentum vectors of the inner and outer discs is between 0 and 2$i$ where $i$ is the angle between the planet's orbital vector and the angular momentum vectors of either the inner or outer disc (the inner and outer discs have the same $i$ initially and the value of $i$ is a constant during the inner disc's nodal precession). Thus, if there is an observed system with a misalignment angle of $i_{obs}$ between the inner and outer discs, it indicates that the orbital plane of the potential planet is misaligned with either the inner or outer discs at an angle of $i_p\ge i_{obs}/2$.  

The minimum planet mass to break the disc is shown in Figure \ref{fig:precessionmass}. This mass increases with the planet's inclination 
since a more inclined planet induces a shallower gap and also a slower precession of the inner disc.
Observationally, if we find a broken disc but cannot constrain the misalignment angle between the inner and outer discs, the minimum planet mass is thus the mass at the $i=0^o$ limit in Figure \ref{fig:precessionmass}.

We can derive this disc-breaking planet mass at the $i=0^o$ limit analytically since the elliptic integral ($K$) at $i=0^o$ is simply one. Combing the gap depth equation at $i=0^o$ (Equation \ref{eq:gapdepth}) with Equations \ref{eq:precessionrate}  and \ref{eq:breaking2}, we can derive a quadratic equation for $q^2$ and the solution for $q$ is shown in Figure \ref{fig:breakingmass}. For a thin disc, a lower mass planet can break the disc and the gap depth is deeper. 
Since these gaps are normally very deep, Equation \ref{eq:gapdepth} can be simplified to $\Sigma_{gap}/\Sigma_0=25/K$. Then, the minimum planet mass to break the disc at the small $i$ limit can be further simplified to
\begin{equation}
    q>\sqrt{20/3}\alpha^{1/4}h^{7/4}\,.\label{eq:minimumq}
\end{equation}
This simple formula agrees with Figure \ref{fig:breakingmass} very well. 

\subsection{Individual Sources}

Here we can use our disc breaking condition (Equations \ref{eq:breaking2} and \ref{eq:minimumq}, Figures \ref{fig:precessionmass} and \ref{fig:breakingmass}) to study the known protoplanetary discs that have clear signatures of disc breaking. 

HD 142527 is a large transitional disc around a 2 $M_{\odot}$ central star \citep{casassus2013}. It has a large cavity ($\sim$140 au) and a noticeable asymmetric structure at the cavity edge. It also has a clear dark lane at its outer disc, which reveals that the inner disc is $\sim$70$^o$ misaligned with the outer disc \citep{Marino2015}. The dark lane coincides with the (sub-)mm intensity dimming inside the asymmetric structure at 140 au \citep{casassus2013}. This dimming is consistent with Figure \ref{fig:scatvortex} where an asymmetric vortex is passing through the inner disc's shadow and the  temperature of the disc under the shadow is decreased. HD 142527 has a known 0.2-0.4 $M_\odot$ companion at 13 au \citep{Biller2012,christiaens2018}. Such a high mass companion is fully capable of breaking the disc (Figures \ref{fig:precessionmass} and \ref{fig:breakingmass}) and causing differential precession between the inner and outer discs.  The companion's orbit can be very eccentric \citep{Lacour2016} and it can explain the observed large cavity \citep{Price2018}. If we assume that the companion's eccentricity is low, the inner disc will precess for a full $2\pi$ angle at a timescale of $\sim$27 planetary orbits (Equation \ref{eq:precessionrate}), which is $\sim$ 900 yrs with a companion at 13 au. Thus over decades, we may see the change of the disc shadow at the outer disc. Here we have used Equation \ref{eq:precessionrate} which has assumed that the inner disc size ($R_d$) is the same as
the planet's semi-major axis ($R_p$). If $R_d$ is much smaller than $R_p$, the precession timescale may be much longer. 

HD 100453 is a protoplanetary disc with two prominent spiral arms \citep{Wagner2015}. The two spiral arms could be excited by the 0.2 $M_{\odot}$ companion (HD 100453B) which is 108 au away from HD 100453A \citep{Dong2016, Wagner2018}. On the other hand, this disc also shows dark lanes in the near-IR scattered light images, indicating a misaligned inner disc \citep{Benisty2017}. The misalignment angle between the inner and outer discs is $\sim$ 72$^o$ and the gap between the inner and outer discs is from 1 to 20 au. One explanation for the misalignment could be that there is a massive inclined planet inside the gap (at several au) and the planet mass can be constrained by Equation \ref{eq:minimumq}. On the other hand, if HD 100453B at 108 au is misaligned with the disc plane of HD 100453A, HD 100453B can also cause precession for both the disc within 1 au and the disc beyond 20 au. Since these two discs have different precession rates, it can lead to misalignment between them. Under this second scenario, massive planets are not needed at several au but HD 100453B should be more than 36$^o$ misaligned with either the inner or outer discs and we should not see the movement of the shadow during decades of observations.
Using Equation \ref{eq:wrbfull} with $R_d$=1 au \citep{Benisty2017}, we can derive $\omega_{rb}/\Omega_b\sim 4\times10^{-5}$ where $\Omega_b$ is the orbital frequency for HD 100453B. Equation \ref{eq:breaking2} then suggests that $\Sigma_{gap}/\Sigma_{0}$ is $\lesssim3\times10^{-6}$ with $h$=0.05.  This scenario may also have applications to other systems with both spirals (which may indicate that there is a massive outer companion) and disc shadows.

DoAr 44 is a protoplanetary disc around a  1.4 $M_{\odot}$ star \citep{Bouvier1992}. It also has a dark lane at the outer disc and the inferred misalignment between the inner and outer discs is $\sim30^o$ \citep{Casassus2018}. The inner and outer discs are separated by a gap from 5-15 au, based on the detailed modeling for its SED and Near-IR scattered light images. Assuming the disc viscosity $\alpha=10^{-3}$ and $h=0.1$,
the minimum planet mass to break the disc is $q\gtrsim$8$\times10^{-3}$ (Equation \ref{eq:minimumq}). The gap depth is thus $\Sigma_{gap}/\Sigma_0\lesssim 4\times10^{-3} $ (Equation \ref{eq:gapdepth}). This planet should be more than $15^o$ inclined with respect to both the inner and outer discs. The inner disc will precess for a full $2\pi$ angle at a timescale of $\sim$300 planetary orbits (Equation \ref{eq:precessionrate}), which is $\sim$ 6000 yrs with a planet at 8 au. Thus,
it will be difficult to detect the position change of the dark lane in the near future, unless the inner and outer discs are oriented in a way which makes $d\theta/dt\gg d\gamma/dt$ (Equation \ref{eq:dthetadt}).

HD 143006 is a protoplanetary disc with gaps. Near-IR scattered light observations reveal asymmetry at the outer disc \citep{Benisty2018}: half of the disc is fainter than the other half. Such asymmetry is consistent with the shadow cast by a mildly inclined inner disc \citep{Facchini2018}.  Assuming that a planet at $\sim$20 au is  responsible for the tilt of the inner disc, Figure \ref{fig:precessionmass} and Equation \ref{eq:minimumq} suggest that the planet mass needs to be larger than 2 $M_J$ in an $\alpha=10^{-3}$ and $H/R=0.04$ disc. 

AA Tau shows signatures of a misaligned inner disc in both the dust continuum image and molecular line channel maps \citep{Loomis2017}.
If the misalignment between the rings is caused by a misaligned planet, the planet needs to be at least in the Jupiter mass range (Equation \ref{eq:minimumq} and Figure \ref{fig:precessionmass}). AA tau
also exhibits a sudden  and long-lasting dimming event at 2011 \citep{Bouvier2013} after 20 years of a constant brightness. If we assume that 20-200 years is the timescale for the inner disc's precession due to a potential companion, the companion is massive and close to the central star (Equation \ref{eq:precessionrate} suggests that the companion needs to be within several au  if $q=0.1$). 

Overall, in order for a misaligned companion to break the disc and lead to differential precession between the inner and outer discs, the gap needs to be relatively deep and a massive giant planet ($q\gtrsim0.001$) is needed. Thus, under this scenario, disc misalignment and shadowing should always be accompanied by deep and wide gaps (e.g. in Pre/Transitional discs), which is consistent with recent observational constraints \citep{Garufi2018}. On the other hand, if we can constrain that the companion mass in the gap of the misaligned disc is small (e.g. $\lesssim M_J$), the disc viscosity has to be very low (e.g. $\alpha\lesssim10^{-4}$) for the disc to break.

\subsection{Limitations And Comparison With Previous Works}
One caveat in this work is that we do not consider the disc's gravitational force on the planet and disc's self-gravity. This excludes the possibility that the planet and disc might undergo secular resonances \citep{lubow2016,OwenLai2017,MatsakosKonigl2017} to generate large misalignment. On the other hand, secular resonances require a massive companion $\sim0.01-0.1 M_{\odot}$ at $\sim$10-100 au from the star and the circumbinary disc is also massive which dominates the angular momentum budget of the system \citep{OwenLai2017}. Thus, secular resonances may work for HD 142527 which has a known massive companion, but it may not work for `Pre/Transitional discs' where we have not found such massive companions in the disc. 

Our work also has implications to secular resonances. With the presence of such a massive planet in the disc to trigger secular resonances, disc breaking can naturally occur (Figure \ref{fig:precessionmass}) and previous analytical studies which separate the system into the circumprimary disc, companion, and circumbinary disc can be justified. 

We also limit the planet's inclination to be less than 39$^o$, so that the inner disc is not undergoing the Kozai-Lidov oscillation \citep{Martin2014}. If the disc is undergoing the Kozai-Lidov oscillation, the excited eccentric disc can be damped by hydrodynamical processes and the disc will settle to a state with an inclination less than $39^o$ and a low eccentricity \citep{Fu2015}. On the other hand, HD 142527 and HD 100453 may be undergoing the Kozai-Lidov oscillation now. These systems have disc misalignment around 70$^o$, implying that the potential planet has a high chance to be misaligned with the inner disc at more than 40$^o$ to trigger the Kozai-Lidov oscillation.

We do not couple radiative transfer calculations with hydrodynamical simulations. Instead, we post-process the hydrodynamical simulations
with Monte-Carlo calculations. This implicitly assumes that the disc has an infinite amount of time to respond to the stellar radiation. This may not be true in realistic protoplanetary discs,  and including both the radiative transfer and the hydrodynamics simultaneously may lead to interesting phenomenon such as generating spiral waves \citep{Montesinos2016}. 

\cite{Arzamasskiy2018} found that, if a planet cannot induce a gap in the disc, the planet can most efficiently warp the disc when the planet's inclination angle is $\sim$2-3 $h$ with respect to the disc. Then, the inclination of the disc warp can reach $\sim 4 q$ (their Figure 8). \cite{Arzamasskiy2018} have also carried out Monte-Carlo radiative transfer calculations for an inclined Jupiter mass planet in a $h=0.1$ disc and no disc shadowing is observed. This is expected because the planet mass just reaches the thermal mass and a shallow gap is induced by the planet. In order to break a $h/r=0.1$ disc with a $q=10^{-3}$ planet, Equation \ref{eq:breaking2} suggests that $\Sigma_{gap}/\Sigma_0<6\times10^{-5}$, which requires a very small numerical viscosity and a long simulation time. Since the planet can not break the disc, the planet induces a warp with the $4q$ disc inclination (0.004 in radians or 0.3$^o$ with $q$=0.001) which is too small to cause any disc shadow. 

Recently, \cite{Nealon2018}  have carried out simulations to show that a Jupiter mass planet can cause a $10^{-3}-10^{-2}$ misalignment between the inner and outer discs while a 6.5 Jupiter mass planet can cause a  $10^{-2}-10^{-1}$ misalignment between the inner and outer discs. Such warp is roughly consistent with the angle of $4q$ derived in \cite{Arzamasskiy2018}. Thus, we should not expect significant disc shadowing from simulations in \cite{Nealon2018} ($0.1$ in radians is only 6$^o$). Disc breaking has not been observed in \cite{Nealon2018} which shows some constant misalignment angle between the inner and outer discs. With a more massive planet (like our simulation \texttt{THINP10I39AM3}), the disc may start to break. 

\section{Conclusion}
Motivated by recent observations that many protoplanetary discs with gaps and holes show signs of disc warping and breaking, we study the interaction between a misaligned massive planet and the protoplanetary disc.   

\begin{itemize}
    \item We analytically derive the relationship between the mass of the misaligned planet and the induced gap depth/width (Equations \ref{eq:depthwidth} and \ref{eq:newK}), by replacing the planet mass with an effective planet mass (Equation \ref{eq:newq}). 
    
    \item We derive the migration and inclination damping rates for misaligned planets that could induce gaps in discs (Equations \ref{eq:ptr} to \ref{eq:tincgap}). 
    
    \item We provide the new disc breaking condition for a disc having a gap (Equations \ref{eq:breaking} and \ref{eq:breaking2}): the sound crossing time through the inner disc is longer than the product of the precession timescale and the square root of the gap depth.
    
    \item Using 3-D hydrodynamical simulations with mesh-refinement, we carry out a series of simulations to study the interaction between the misaligned planet and the disc. We  find that 10 grid points per scale height is needed for studying the disc secular evolution. A higher order reconstruction scheme (e.g. PPM) can lower this requirement to 5 grid points per scale height. 
    
    \item The simulations show that when the gap is deep enough, the inner and outer discs precess independently so that the disc breaks. Even after the disc breaking, the circumplanetary disc around the planet seems to keep its alignment with the outer disc instead of the inner disc. 
    
    \item If a gap is not induced, the planet's migration and inclination damping rates are similar to the previous study for low mass planets. When a gap is induced, both rates decrease as the gap gets deeper. The simulation results are consistent with analytical expectations (Equations \ref{eq:ptr} to \ref{eq:tincgap}) as long as the disc does not break.
    
    \item The induced gap depth from simulations is consistent with the analytical expectation very well (Equations \ref{eq:depthwidth} and \ref{eq:newK}). For simulations that show disc breaking,
    our derived disc breaking condition (Equation \ref{eq:breaking}, Figure \ref{fig:gapbreaking}) also agrees with simulation results very well.
    
    \item We  generate near-IR scattered light images and (sub-)mm CO velocity channel maps for broken discs, by importing the simulated broken disc structure into Monte-Carlo radiative transfer calculations. In the near-IR scattered light images for inviscid simulations, we see that the vortex at the outer gap edge is traveling through the shadow cast by the inner disc. When the vortex and the shadow overlaps, the vortex looks like it splits into two vortices. The temperature of the disc region that is under the shadow decreases, potentially leading to a change of the (sub-)mm intensity there. 
    
    \item The disc shadow in near-IR scattered light images is also affected by the disc surface density. A higher surface density means that the scattering surface is higher and the shadow reveals this bow-shaped flaring surface.
    
    \item We  observe that the shadow rotates at a different orbital frequency from the precession frequency of the inner disc. The relationship between these two frequencies is derived (Equation \ref{eq:dthetadt}).
    
    \item CO channel maps clearly reveal the geometry of the broken disc. If the molecular line is optically thick within the gap, the channel maps show a twisted velocity structure. 
    
    \item If observations  reveal the broken disc geometry, we can use this geometry to constrain the planet mass (Figures \ref{fig:precessionmass} and \ref{fig:breakingmass}) and the inclination. The minimum planet mass which can break the disc is derived analytically (Equation \ref{eq:minimumq}). We  use this criterion to explore the potential companions in HD 100453, DoAr 44, HD 143006, and AA Tau. 
    
     \item On the other hand, if we can constrain that the companion mass in the gap of the misaligned disc is small (e.g. $\lesssim M_J$), the disc viscosity has to be very low (e.g. $\alpha\lesssim10^{-4}$) for the disc to break. This provides some indirect constraints on the disc viscosity. 
    
    \item Overall, in order for a misaligned companion to break the disc and lead to differential precession between the inner and outer discs, the gap needs to be relatively deep (e.g. in Pre/Transitional discs), and a planet with at least a Jupiter mass is needed even under some extreme disc conditions. Our study supports the scenario that massive planets are present in Pre/Transitional discs that have disk shadows, where the massive planets (probably larger than $M_J$) are responsible for both the wide gaps/cavities and the inner disc misalignment causing disc shadowing. These disks with shadows are thus prime targets for directly searching exoplanets in protoplanetary discs.

\end{itemize}

\section*{Acknowledgements}
Z. Z. thank Jaehan Bae for sharing his scripts to generate CO channel maps.
Z. Z. thank Steve Lubow for very helpful discussions. 
Z. Z. acknowledges support from the National Aeronautics and Space Administration 
through the Astrophysics Theory Program with Grant No. NNX17AK40G and Sloan Research Fellowship. 
Simulations are carried out with the support from the Texas Advanced Computing Center (TACC)
at The University of Texas Austin through XSEDE grant TG-AST130002, and 
the NASA High-End Computing (HEC) Program through the NASA Advanced Supercomputing (NAS)
Division at Ames Research Center.

\input{paper.bbl}

\bsp
\label{lastpage}
\end{document}